\documentclass[fleqn,usenatbib,useAMS]{mnras}

\usepackage{graphicx}	
\usepackage{amsmath}
\usepackage{amssymb}	
\usepackage{bm}
\usepackage{physics} 
\usepackage{multicol}

\newcommand{\p}{\partial}
\newcommand{\OmK}{\Omega_\text{K}}

\newcommand{\rmB}{\mathrm{B}}
\newcommand{\rmR}{\mathrm{R}}
\newcommand{\rmk}{\mathrm{k}}

\newcommand{\rmi}{\mathrm{i}}

\newcommand{\rmv}{\mathrm{v}}
\newcommand{\rmz}{\mathrm{Z}}
\newcommand{\cs}{\mathrm{c}_\mathrm{s}}
\newcommand{\qD}{q_\mathrm{D}}
\newcommand{\qT}{q_\mathrm{T}}
\newcommand{\ik}{\mathrm{i}\mathrm{k}}
\newcommand{\az}{{AZ}}

\newcommand{\bB}{\mathbf{B}}
\newcommand{\bk}{\mathbf{k}}
\newcommand{\bJ}{\mathbf{J}}
\newcommand{\bv}{\mathbf{v}}
\newcommand{\bva}{\mathbf{v_A}}
\newcommand{\nn}{\nonumber}

\title{On the Vertical Shear Instability in Magnetized Protoplanetary Disks}

\author[]{Can Cui$^{1}$\thanks{E-mail: \href{mailto:cc795@cam.ac.uk}{cc795@cam.ac.uk}} and 
Min-Kai Lin$^{2,3}$\thanks{E-mail: \href{mailto:mklin@asiaa.sinica.edu.tw}{mklin@asiaa.sinica.edu.tw}}
\\
$^{1}$DAMTP, University of Cambridge, CMS, Wilberforce Road, Cambridge CB3 0WA, UK \\
$^{2}$Institute of Astronomy and Astrophysics, Academia Sinica, Taipei 10617, Taiwan  \\
$^{3}$Physics Division, National Center for Theoretical Sciences, Taipei 10617, Taiwan}

\pubyear{2021}

\begin{document}
\label{firstpage}
\pagerange{\pageref{firstpage}--\pageref{lastpage}}
\maketitle

\begin{abstract}

The vertical shear instability (VSI) is a robust   phenomenon in irradiated protoplanetary disks (PPDs). While there is   extensive literature on the VSI in the hydrodynamic limit, PPDs are expected to be magnetized and their extremely low ionization fractions imply that non-ideal magneto-hydrodynamic (MHD) effects should be properly considered. To this end, we present linear analyses of the VSI in magnetized disks with Ohmic resistivity. We primarily consider toroidal magnetic fields, which are likely to dominate the field geometry in PPDs. We perform vertically global and radially local analyses to capture characteristic VSI modes with extended vertical structures. To focus on the effect of magnetism, we use a locally isothermal equation of state. We find that magnetism provides a stabilizing effect to dampen the VSI, with surface modes, rather than body modes, being the first to vanish with increasing magnetization. Subdued VSI modes can be revived by Ohmic resistivity, where sufficient magnetic diffusion overcome magnetic stabilization, and hydrodynamic results are recovered. We also briefly consider poloidal fields to account for the magnetorotational instability (MRI), which may develop towards surface layers in  the outer parts of PPDs. The MRI grows efficiently at small radial wavenumbers, in contrast to the VSI. When resistivity is considered, we find the VSI dominates over the MRI for Ohmic Els\"{a}sser numbers $\lesssim 0.09$ at plasma beta parameter $\beta_Z \sim 10^4$. 
\end{abstract}


\section{Introduction}
It has been postulated over decades that the turbulence and angular momentum transport in most astrophysical accretion disks are mediated by the magnetorotational  instability (MRI; \citealp{bh91}). However, protoplanetary disks (PPDs) are distinguished by their extremely weakly ionized gas \citep{gammie96,armitage11}, where gas and magnetic fields are poorly coupled, and the MRI turbulence is either quenched or dampened in the bulk of the disk \citep{pc11,bs11,simon_etal13a,simon_etal13b}. Instead, angular momentum transport is dominated by magnetized disk winds, leaving the main disk mostly laminar \citep{bs13,gressel_etal15,bai+16,bai17,bethune_etal17,gressel_etal20}. 

Nevertheless, some level of turbulence is expected in PPDs to account for the recent ALMA observations of molecular line emissions \citep{teague_etal16,flaherty_etal17,flaherty_etal18,flaherty_etal20}. Furthermore, turbulence may serve as an essential ingredient in many stages of planet formation. Turbulence affects the gravitational sedimentation \citep{dubrulle_etal95,jk05,yl07}, radial diffusion \citep{cp88}, and collisional growth \citep{oc07,birnstiel_etal10} of dust particles. Long-lived vortices induced by turbulence (e.g. \citealp{raettig_etal15,mk18}) can concentrate dust particles \citep{bs95,kh97,cuzzi_etal08} and seed planetesimal formation through streaming instability or gravitational instability \citep{yg05,johansen_etal07,cy10}, whereas the growth of streaming instability can be substantially diminished by a moderate level of turbulent viscosity \citep{cl20,umurhan20}. Turbulence also influences the radial migration of planets, as well as the flow morphology and gap formation around them \citep{np04,pn04}. Therefore, understanding the origin and characteristics of turbulence in PPDs is essential to many aspects of planet formation and evolution.

The lack of MHD turbulence in a PPD led to a surge in the interest of purely hydrodynamic instabilities \citep{lu19,weiss_etal21}. Among the most explored are the vertical shear instability (VSI; \citealp{nelson_etal13}, hereafter N13, \citealp{ly15,lp18}, hereafter LP18, \citealp{cb20}), the convective overstability in its linear \citep{kh14,lyra14,latter16} and non-linear (subcritical baroclinic instability; \citealp{kb03,petersen_etal07a,petersen_etal07b,lp10}) phases, and the zombie vortex instability \citep{marcus_etal13,marcus_etal15,umurhan_etal16a,ll16}. These instabilities set in under certain thermodynamic and structural conditions, thereby operating at distinct regions of PPDs \citep{malygin_etal17,pk19,lu19}. The VSI is of particular interest as it extends a large portion of the disk (e.g. \citealp{sk14,ly15,flock_etal20,pk20}). 

The VSI is inherited from the Goldreich--Schubert--Fricke instability \citep{gs67,fricke68} and is initially discovered in the context of differentially rotating stars. Its importance to accretion disks was later explored by \citet{ub98}, \citet{urpin03} and \citet{au04}. The applicability of the VSI to PPDs has been demonstrated only recently in \citealp{nelson_etal13}. It quickly drew intensive interests (e.g.  \citealp{sk14,bl15,umurhan_etal16b,lp18,lin19,cb20,schafer_etal20}) and is considered to be a promising hydrodynamic mechanism in driving turbulence in PPDs.
 
A differentially rotating disk with Keplerian profile is stable according to the Rayleigh criterion \citep{chandrasekhar61}. The presence of vertical shear can destabilize inertial waves in a vertically global disk model (\citealp{bl15}). Nevertheless, a fluid element also experiences stabilizing effects from vertical buoyancy, which impedes the VSI growth. This can be overcome by sufficiently rapid cooling that brings the perturbed fluid element to reach local thermal equilibrium with its surroundings, hence diminishing the buoyancy. Local linear analyses demonstrate that the unstable modes are characterized by short radial wavelengths and maximum growth rates much smaller than the orbital frequency (\citealp{urpin03}; N13). Two classes of VSI modes have been identified: rapidly-growing surface modes concentrated near the disk surface and more vertically extended body modes (N13; \citealp{bl15,ly15}). The body modes can be further categorized into breathing and corrugation modes, depending on the symmetry about the midplane (N13). 
 
The non-linear evolution of the VSI has been examined by  hydrodynamic simulations. In accordance with the linear  theory, the VSI is triggered when thermal relaxation  timescales are less than $0.01-0.1$ times the local dynamical timescales, and the wave modes exhibit elongated vertical wavelengths (N13). While the surface modes possess the fastest growth rate, the body (corrugation) modes eventually take over, dominating the non-linear evolution. Fully developed VSI turbulence yield a Shakura–Sunyaev \citep{ss73} $\alpha$ value on the order of $10^{-4} - 10^{-3}$ (N13, \citealt{sk14}). Non-axisymmetric 3D simulations show the development of vortices  \citep{richard_etal16,mk18,flock_etal20,manger_etal20,pk20}. Incorporating dust particles, numerical simulations show that VSI can stir up dust grains against vertical settling, but may also concentrate them through inducing dust-trapping vortices \citep{sk16,flock_etal17,lin19,schafer_etal20}. 

Most studies of the VSI to date have neglected magnetic fields despite their importance in the evolution of PPDs. Two recent works extend analyses of the VSI to MHD regimes. Local linear stability analyses in the ideal MHD limit show that weak magnetic fields are favoured to excite the VSI, specifically when plasma beta $\beta \gtrsim 400$ for thin disks (LP18), where $\beta$ is the ratio of gas to magnetic pressure. They also find the MRI growth rates exceed that for VSI modes, but the wave vectors of the two are perpendicular to each other. For resistive disks, LP18 estimates a critical Ohmic Els\"{a}sser number of $\sim 1/h$, where $h$ is the disk aspect ratio, for the VSI to operate, by requiring the magnetic diffusion timescale to be shorter than the Alfv\'en wave propagation timescales. Non-linear MHD simulations, applicable to outer regions of the disk, demonstrate that the VSI can initiate and sustain turbulence in magnetized disks \citep{cb20}. Weak ambipolar diffusion strength, or the enhanced coupling between gas and magnetic fields, works as stabilizing effects to dampen the VSI growth. 

Previous hydrodynamic models suggest that effective VSI growths span over $\sim5-100$ AU in PPDs \citep{ly15,pk19}. These regions are susceptible to all three non-ideal MHD effects -- Ohmic resistivity, Hall effect, and ambipolar diffusion \citep{wardle07,bai11}. However, a quantitative analysis in vertically-global, magnetized disks with non-ideal effects is still lacking. Such analyses can be useful for understanding the VSI mode properties in real PPDs and interpreting nonlinear simulations. Hence, in this work, we extend the linear stability analysis of the VSI to weakly ionized gas in a vertically global disk model. We remark that a vertically-global analysis is necessary for a proper description of elongated body modes of the VSI, which have been found to dominate in numerical simulations (N13; \citealp{sk14,cb20}). We consider ideal MHD and further include Ohmic resistivity as a proxy for non-ideal effects. We focus primarily on the effect of toroidal magnetic fields. However, in a local model, we also investigate the dominance between MRI and VSI by considering purely poloidal magnetic fields. 

The plan of the paper is as follows: in \S\ref{sec:be}, we introduce the basic formulation and establish the equilibrium state of the problem. 
In \S\ref{sec:SH}, we derive and discuss the Solberg-Hoiland stability criteria for magetized disks. In \S\ref{sec:lp}, we present the linearized equations and detail the analytical and numerical methods used, with results shown in \S\ref{sec:re}, for a purely toroidal background magnetic field. We also conduct a brief analysis for a purely poloidal background magnetic field and compare the MRI growth rates with the VSI in \S\ref{sec:po}. Finally, we discuss the results in \S\ref{sec:di} and summarize our main findings in \S\ref{sec:sum}. 

\section{Basic Equations and Equilibria}\label{sec:be}
Consider a gaseous, inviscid, magnetized PPD. The gas density, velocity, and magnetic field are denoted by $\rho$, $\bv$, $\bB$, respectively. Our formulations are presented in cylindrical $(R,\phi, \rmz)$ coordinates centered on the protostar, although the spherical radius $r$ is also used to simplify expressions. The basic dynamical equations written in SI units are 
\begin{equation}
\dv{\rho}{t} + \rho \nabla \cdot \bv = 0, 
\label{eq:cont}
\end{equation}
\begin{equation}
\rho\dv{\bv}{t} + \nabla \Pi- \frac{1}{\mu_0}(\bB\cdot\nabla)\bB + \rho\nabla\Phi=0, 
\label{eq:eom}
\end{equation}
where ${\rm d/d}t \equiv \p/\p t + \bv \cdot\nabla$ is the material derivative and $\mu_0$ is the magnetic permeability. The gravitational potential is given by 
\begin{equation}
\Phi=-\frac{GM}{r}, 
\end{equation}
where $G$ is gravitational constant, and $M$ is the mass of the central star. 
The total pressure 
\begin{align}\label{ptot_def}
\Pi= P+P_\rmB
\end{align}
is the sum of thermal pressure $P$ and magnetic pressure 
\begin{align}
P_\rmB = \frac{\rmB^2}{2\mu_0}, \label{pmag_def}
\end{align}
where $\rmB = |\bB|$. The strength of the magnetic field is parametrized by the ratio of gas pressure to magnetic pressure,
\begin{equation}
\beta = \frac{P}{P_\rmB}.
\label{beta_def}
\end{equation} 

\subsection{Induction Equation}\label{sec:idc}
The evolution of magnetic fields is governed by the induction equation, in which three non-ideal MHD effects manifest,
\begin{equation}
\frac{\partial \bB}{\partial t}  = \nabla \times \bigg[ \bv \times \bB - \frac{\bJ}{\sigma_B}  - \frac{\bJ \times \bB}{en_e}  + \frac{(\bJ \times \bB) \times \bB}{\gamma_i \rho \rho_i} \bigg].
\label{eq:idc}
\end{equation}
Here, $\sigma_B$ is resistivity, $e$ is electric charge, and $n_e$ is electron number density. The drag coefficient $\gamma_i$ represents the momentum transfer between ion-neutral collisions, and the ion density is denoted by $\rho_i$. On the right-hand side, the four terms each corresponds to the standard inductive term, Ohmic resistivity, the Hall effect, and ambipolar diffusion. Ohmic resistivity is relevant to electron-neutral collisions. Ambipolar diffusion corresponds to ion-neutral drift. The Hall effect is associated with ion-electron drift, and it differs fundamentally from Ohmic resistivity and ambipolar diffusion as being a non-dissipative process. 
The relation between the current density $\bJ$ and the magnetic field is completed by the Maxwell equation, 
\begin{equation}
\bJ = \frac{1}{\mu_0} \nabla \times \bB.
\label{eq:J}
\end{equation}
The displacement current in Equation \eqref{eq:J} is self-consistently neglected for non-relativistic MHD \citep{balbus09}. Finally, the magnetic field also satisfies the solenoidal condition, 
\begin{equation}
\nabla\cdot\bB=0.
\end{equation}

To characterize the non-ideal MHD effects, it is convenient to define the diffusion coefficients 
\begin{equation}
\eta_O = \frac{1}{\mu_0\sigma_B}, \quad \eta_H = \frac{\rmB}{\mu_0e n_e}, \quad \eta_A=\frac{\rmB^2}{\mu_0\gamma_i \rho \rho_i},
\end{equation}
for Ohmic resistivity, the Hall effect, and ambipolar diffusion. At a given ionization fraction, the Ohmic diffusivity $\eta_O$ is independent of field strength and density, whereas the Hall diffusivity $\eta_H\propto B/\rho$, and the ambipolar diffusivity $\eta_A\propto B^2/\rho^2$. Hence, ambipolar diffusion dominates in regions of strong fields or low densities, Ohmic resistivity dominates in regions of weak fields or high densities, and the Hall effect governs in between \citep{lesur20}.
We further introduce dimensionless Els\"{a}sser numbers,  
\begin{equation}
\Lambda=\frac{\rm\rmv_\textrm{A}^2}{\eta_O\OmK}, \quad \chi=\frac{\rm\rmv_\mathrm{A}^2}{\eta_\mathrm{H}\OmK}, \quad {\rm Am}=\frac{\rm\rmv_\mathrm{A}^2}{\eta_\mathrm{A}\OmK}
\label{eq:elsasser}
\end{equation}
where the Alfv\'{e}n velocity $\rm\rmv_{\mathrm A}$ is given by
\begin{equation}
\rmv^2_\mathrm{A}=\frac{|\bB|^2}{\mu_0 \rho}. 
\end{equation}

\subsubsection{Toroidal fields in axisymmetric disks}\label{sec:toraxis}

Toroidal magnetic fields have been shown to dominate
in global non-ideal MHD simulations, with the saturation of VSI turbulence (e.g \citealp{bethune_etal17,bai17,cb20}). Furthermore, the VSI is an axisymmetirc instability (N13). 
These results motivate us to primarily consider purely toroidal fields in axisymmetric disks. 
In this case, the Hall effect vanishes, and we show here that ambipolar diffusion behaves the same way as Ohmic resistivity. In the limit of $\bB= (0, B_\phi, 0)$, we expand the numerator of the last term on the right-hand side of Equation \eqref{eq:idc}, corresponding to ambipolar diffusion,
\begin{equation}
(\bJ \times  \bB) \times \bB = (\bJ \cdot \bB) \bB - \rmB^2\bJ.
\label{eq:ad}
\end{equation}
This implies that for electric currents being everywhere perpendicular to magnetic fields, as the case for purely toroidal field geometry, ambipolar diffusion can be treated as an effective resistivity with field dependency \citep{bt01}. 
Consequently, the induction equation reduces to 
\begin{equation}
\dv{\rm\rmB_\phi}{t} =  - (\nabla \cdot\bv)\rmB_\phi + \frac{\rm\rmB_\phi \rm\rmv_\rmR}{R} + \eta \left( \nabla^2\rm\rmB_\phi - \frac{B_\phi}{R^2}\right),
\label{bphi_evol}
\end{equation} 
where $\eta = \eta_O + \eta_\mathrm{A}$, and we assume a constant $\eta$ throughout. The second and final terms on the right-hand side result from the curvilinear geometry.  

\subsection{Effective Energy Equation}\label{sec:effectiveE}
The ideal gas law is given by 
\begin{equation}\label{ideal_gas}
P = \frac{\mathcal{R}}{\mu}\rho T,
\end{equation}
where $\mathcal{R}$ is the gas constant, $\mu$ is the mean molecular weight, and $T$ is the gas temperature.
We define the isothermal sound speed $c_s$, pressure scale-height $H$, and disk aspect-ratio  $h$ through
\begin{equation}
\cs^2 \equiv \frac{P}{\rho}, \quad H \equiv \frac{\cs}{\OmK}, \quad h\equiv \frac{H}{R},
\end{equation}
where $\OmK = \sqrt{GM/R^3}$ is the Keplerian angular velocity.
In the hydrodynamic limit, isothermality is most favourable for the VSI because the stabilizing effect stabilizing from gas buoyancy is absent (N13; \citealp{ly15}). 
To focus on the effect of magnetic fields, we adopt a locally isothermal thermodynamic response, which is applicable to the outer parts of PPDs wherein the temperature is regulated by stellar irradiation \citep{chiang97}. 
Thereby, the energy equation of gas pressure becomes 
\begin{equation}
\dv{P}{t} + P \nabla \cdot \bv = \rho\bv\cdot\nabla \cs^2.
\label{eq:gpress}
\end{equation} 

Consider the evolution of the magnetic pressure associated with a purely azimuthal field, 
\begin{align}
\dv{P_\rmB}{t} = \frac{\rm\rmB_\phi}{\mu_0} \dv{\rm\rmB_\phi}{t}. 
\label{eq:pb}
\end{align}  
Combining Equations \eqref{bphi_evol}, \eqref{eq:gpress}, and \eqref{eq:pb}, we formulate an effective energy equation 
\begin{align}
\dv{\Pi}{t}  + \gamma_\rmB \Pi(\nabla \cdot \bv) & = \rho\bv\cdot\nabla \cs^2 + \frac{B_\phi^2}{\mu_0R}v_R \notag \\
&+ \frac{\rmB_\phi}{\mu_0} \eta \left(\nabla^2 \rmB_\phi - \frac{B_\phi}{R^2}\right),
\label{eq:energy}
\end{align} 
where the effective adiabatic index is defined by 
\begin{equation}
\gamma_\rmB \equiv \frac{\beta+2}{\beta+1}. 
\end{equation}
Equation \eqref{eq:energy} resembles the energy equation for a fluid subject to heating or cooling in the hydrodynamic limit \citep{ly15}. It shows that an axisymmetric, locally isothermal, magnetized gas with a purely azimuthal field behaves like an unmagnetized gas with adiabatic index $\gamma_\rmB$\footnote{We remark that this analogy can be generalized to locally polytropic disks where $P=K\rho^\Gamma$, where $K$ is a prescribed function of position and $\Gamma$ is the polytropic index. In this case the effective adiabatic index becomes $\gamma_B= (\Gamma\beta + 2)/(\beta + 1)$.}, while it is also subject to non-adiabatic effects on the right-hand side: the locally isothermal thermodynamic response, magnetic diffusion, and curvature effects. Note that $\gamma_\rmB(\beta)$ here is not necessarily constant because $\beta(R, Z, t)$ can be non-uniform and can evolve in time. For $\beta \to \infty$, we recover a locally isothermal gas with unit adiabatic index $\gamma_\rmB \to 1$. On the other hand, a strongly magnetized disk with $\beta\to 0$ is equivalent to a fluid with an effective adiabatic index $\gamma_\rmB \to 2$. 

\subsection{Equilibrium State}\label{sec:eqmdisk}
We consider axisymmetric steady states with a purely azimuthal velocity and magnetic field. 
The midplane gas density and temperature are prescribed as
\begin{equation}
\rho_{\rm 0} (R) = \rho_{\rm 0} (R_0)\bigg(\frac{R}{R_0}\bigg)^{-\qD}, 
\end{equation} 
\begin{equation}
T (R) = T(R_0) \bigg(\frac{R}{R_0}\bigg)^{-\qT}, 
\end{equation} 
where $R_0$ is a reference radius, and $\qT$ and $\qD$ are constant power-law indices. 
The equilibrium solutions satisfy the equation of motion,
\begin{equation}
\frac{\partial \Phi}{\partial R} + \frac{1}{\rho}\frac{\partial \Pi}{\partial R} + \frac{2P_\rmB}{\rho R} = R\Omega^2,  \label{eq:eomR} 
\end{equation} 
\begin{equation}
\frac{\partial \Phi}{\partial Z} + \frac{1}{\rho}\frac{\partial \Pi}{\partial Z}=0,
\label{eq:eomz}
\end{equation} 
where $\Omega\equiv v_\phi/R$ is the angular frequency. These equations may be solved explicitly for a given  magnetic field configuration. We consider two special cases in the following, depending on whether resistivity is included. 

\subsubsection{Constant-$\beta$ disks} \label{sec:cbeta}
We consider constant-$\beta$ disks for ideal MHD ($\eta=0$). The effective energy equation \eqref{eq:energy} is then satisfied identically so the solutions below represent exact equilibria.  
In this case, the vertical gradient of equilibrium angular velocity is
\begin{equation}
\frac{\partial \Omega^2}{\partial Z} =  - \frac{\qT}{R^2} \frac{GMZ}{r^3}.
\label{eq:shear}
\end{equation} 
The density and rotation profiles are 
\begin{equation}
\rho (R,Z)  = \rho_0(R)\exp[h^{-2} \left(\frac{R}{r}-1\right) \frac{\beta}{1+\beta}],
\label{rhoRz}
\end{equation} 
\begin{equation}
\Omega^2 (R,Z) = \Omega_{\rm K}^2 \bigg[(1-\qT) -  \frac{1+\beta}{\beta} \bigg(\qT+\qD + \frac{2}{1+\beta}\bigg)h^2 +  \frac{\qT R}{r} 
\bigg].
\label{eq:omg}
\end{equation} 
As $\beta\to\infty$, we recover the hydrodynamic limit (N13). The thin disk approximation ($h \ll 1$) leads to $\Omega \approx \OmK$. As $\beta\to 0$, the density gradient profile shows that a strongly magnetized disk becomes vertically unstratified. A small $\beta \leqslant 1$ leads to strong deviations from the Keplerian rotation, but a minimum value of $\beta$ is required to ensure $\Omega^2>0$ at the disk midplane,
\begin{equation}
\beta > \frac{\left(\qT + \qD - 2\right)h^2}{ 1 - \left(\qT + \qD\right)h^2}.
\end{equation} 
For typical PPD parameters, $h\sim 0.05$ and $\qD,\, \qT$ are of order unity, so the above requirement becomes $\beta\gtrsim O(10^{-3})$. This is easily satisfied for the weakly magnetized disks with $\beta \gtrsim 1$ that we consider. 

\subsubsection{Constant-$B_\phi$ disks}\label{sec:cBphi}

We consider constant-$B_\phi$ disks for ideal MHD and resistive disks ($\eta\neq0$). Resistive disks present approximate equilibrium solutions because there is a slow diffusion of the magnetic field due to the global curvature term in Equation \eqref{eq:energy}. However, 
this is not expected to significantly affect radially localized dynamics such as the VSI.  
In this case, $\beta$ is no longer a constant but declines with height,
\begin{equation}
\beta(R, Z) = \beta_0(R)\frac{\rho(R, Z)}{\rho_0(R)}, 
\label{eq:beta} 
\end{equation}
where $\beta_0 = \beta(R,0)$. The disk becomes more strongly magnetized with increasing height. 

For constant-$B_\phi$ disks, magnetic pressure does not contribute to the vertical equilibrium, and we recover the hydrodynamic limit for density profile
\begin{equation}
\rho (R,Z) = \rho_0(R)\exp[h^{-2} (R/r - 1)]. 
\end{equation} 
The vertical shear gradient differs from constant-$\beta$ disks because the curvature term in Equation \eqref{eq:eomR} now depends on $Z$ through $\rho$. We find
\begin{equation}
\frac{\partial \Omega^2}{\partial Z}  =\left(\frac{2}{\beta}-\qT\right) \frac{1}{R^2} \frac{GMZ}{r^3},
\label{eq:dO2dZ2} 
\end{equation} 
and the rotation profile is
\begin{equation}
\Omega^2 (R,Z) = \Omega_{\rm K}^2  \bigg[(1-\qT) - \bigg(\qT+\qD + \frac{2}{\beta} \bigg)h^2 +\frac{\qT R}{r} \bigg].
\label{eq:shear2} 
\end{equation} 
Accordingly, even a strictly isothermal disk ($\qT=0$) can exhibit vertical shear due to the curvature term. Thus, we may expect a corresponding VSI growth{, but this} should be examined further in radially global models. For PPDs where magnetic fields are weak, the angular velocity profiles in Equations \eqref{eq:omg} and \eqref{eq:shear2} converge to the hydrodynamic limit.  

\section{Solberg-Hoiland Criteria}\label{sec:SH}

The Solberg-Hoiland criteria describe the linear hydrodynamic stability of ideal fluids against axisymmetric and adiabatic perturbations \citep{tassoul78}. Now, our axisymmetric, magnetized disks with purely azimuthal fields obey a similar energy Equation \eqref{eq:energy} as in hydrodynamics. If, in addition, curvature terms and magnetic tension forces can be neglected, then our disk models satisfy the same form of equations as in adiabatic hydrodynamics. This motivates us to formulate an equivalent Solberg-Hoiland stability  criteria for magnetized disks as follows.

The standard hydrodynamic Solberg-Hoiland criteria are expressed in terms of the gradient of pressure $P$, entropy $S\propto \ln{\left(P^{1/\gamma}/\rho\right)}$, and angular frequency $\Omega$, where $\gamma$ is the adiabatic index. By setting $P\to \Pi$ and $\gamma\to\gamma_\rmB$ , we find that the stability is ensured in a magnetized disk if
\begin{equation}
\kappa^2 - \frac{1}{\rho} \nabla \Pi\cdot \nabla S_\rmB > 0, 
\label{SH_mag1}
\end{equation}
\begin{equation}
-\frac{1}{\rho}\frac{\partial \Pi}{\partial Z}\bigg(\kappa^2\frac{\partial S_\rmB}{\partial Z} - R\frac{\partial \Omega^2}{\partial Z} \frac{\partial S_\rmB}{\partial R} \bigg) > 0,
\label{SH_mag2}
\end{equation}
where
\begin{equation}
\kappa^2 \equiv \frac{1}{R^3}\frac{\partial\left(R^4 \Omega^2\right)}{\partial R} \nonumber
\end{equation}
is the square of epicyclic frequency, and
\begin{equation}
\nabla S_\rmB \equiv \frac{1}{\gamma_\rmB} \nabla \ln{\Pi} - \nabla \ln{\rho}
\end{equation}
defines the gradient of effective entropy $S_\rmB$ of a locally isothermal fluid with an azimuthal field. Correspondingly, we define 
\begin{align}\label{Nz_def}
N_\rmR^2 = -\frac{1}{\rho}\pdv{\Pi}{R} \pdv{S_\rmB}{R}, \quad N_\rmz^2 = -\frac{1}{\rho}\frac{\p \Pi}{\p Z} \frac{\p S_\rmB}{\p Z}
\end{align}
as the square of buoyancy frequencies in the radial and vertical directions, respectively.

To examine whether magnetized disk models are stable subject to the modified Solberg-Hoiland criteria, we first derive the vertical buoyancy frequency for constant-$\beta$ and constant-$B_\phi$ thin disks, 
\begin{equation}
N_\rmz^2 = \frac{\beta}{(\beta+2)(\beta+1)}\frac{Z^2}{H^2}\OmK^2 \quad \text{ (constant-$\beta$)},
\label{eq:Nzbeta}
\end{equation}
\begin{equation}
N_\rmz^2 = \frac{2}{\beta(R,Z) + 2}\frac{Z^2}{H^2}\OmK^2 \quad \text{ (constant-$B_\phi$)}.
\label{eq:Nzbphi} 
\end{equation}
As $\beta \rightarrow \infty$, $N_\rmz^2\propto 1/\beta \to 0$ for both models, and we recover a locally isothermal hydrodynamic disk with vanishing vertical buoyancy. As $\beta \rightarrow 0$, for constant-$\beta$ disks, $N_\rmz^2 \propto \beta$ and the buoyancy effect vanishes since there is no vertical density stratification in this limit. In a constant-$B_\phi$ disk, $\beta \rightarrow 0$ occurs for large $|Z|$ as in Equation \eqref{eq:beta}, so that $N_\rmz^2 \to \OmK^2Z^2/H^2$, and the disk expected to be strongly stabilized by magnetic buoyancy. 

We now examine whether the first Solberg-Hoiland criterion is satisfied. Equation \eqref{SH_mag1} can be cast into 
\begin{equation}
\kappa^2 + N_\rmR^2 + N_\rmz^2 > 0. 
\label{eq:SH_mag1_}
\end{equation}
 A stable stratification, $N_{R,Z}^2>0$, requires the effective entropy to increase with $R$ or $Z$ because total pressure gradients are usually negative. In thin weakly magnetized disks, $\left|N_\rmR\right|\sim O(\OmK H/R)\ll \kappa$ and $N_\rmz^2>0$ (Equations \eqref{eq:Nzbeta} and \eqref{eq:Nzbphi}), thus the first criterion is generally satisfied.
 
Next, we examine the second Solberg-Hoiland criterion. Recall that the Solberg-Hoiland criteria apply to adiabatic flows. Thus, Equations \eqref{SH_mag1} and \eqref{SH_mag2} should only be applied if the governing equations of the magnetized disk (Equations \eqref{eq:cont}--\eqref{eq:eom}, \eqref{eq:energy}) can map exactly to adiabatic hydrodynamics. That is, the right-hand side of Equation \eqref{eq:energy} and the magnetic tension force in Equation \eqref{eq:eom} should be negligible. This requires  
\begin{enumerate}
\item a strictly isothermal disk (constant $\cs$ in $R,Z$);
\item ideal MHD ($\eta=0$);  
\item weak magnetic fields ($\beta\gg 1$) so that curvature terms associated with magnetic fields can be neglected \citep{pp05}.
\end{enumerate}
The first restriction implies $\p\Omega/\p Z=0$. The second criterion then becomes 
\begin{align} 
\kappa^2N_\rmz^2 > 0,
\end{align}
which is satisfied in both of our disk models. To study stability with vertical shear, which requires a radially varying disk temperature and non-ideal MHD effects, we must solve the linearized equations explicitly as conducted in the following section.

\section{Linear problem}\label{sec:lp}

\subsection{Perturbation Equations}
We consider axisymmetric Eulerian perturbations for $\bv^\prime, \rho^\prime$, and $\Pi'$ of the form
\begin{equation}
\rho^\prime \propto \exp (\sigma t + \ik_R R).
\end{equation} 
The complex frequency is denoted by $\sigma = s + \rmi \omega$, and a real radial wavenumber $\rmk$ is taken. We assume radially localized disturbances, $\rmk_R R\gg 1$. The background disk variables are evaluated at the reference radius $R_0$, but their vertical dependence is retained. Curvature terms resulting from the cylindrical geometry are neglected, which restricts the analyses to weak magnetic fields \citep{pp05}. By assuming $\bB'=(0,B_\phi',0)$ for simplicity, the linearized perturbation equations read
\begin{equation} 
\sigma \rmv_\rmR' - 2\Omega \rmv_\phi' - \frac{\rho'}{\rho^2} \pdv{\Pi}{R}  + \ik_R \frac{\Pi^{{'}}}{\rho} = 0,  
\label{eq:l1}
\end{equation} 

\begin{equation} 
\sigma \rmv_\phi' + \frac{\kappa^2}{2\Omega} \rmv_\rmR'  + \pdv{\rmv_\phi}{Z} \rmv_\rmz' = 0, 
\end{equation} 

\begin{equation} 
\sigma \rmv_\rmz'   - \frac{\rho'}{\rho^2}\pdv{\Pi}{Z}  + \frac{1}{\rho}\pdv{\Pi'}{Z}  = 0, 
\end{equation} 

\begin{equation} 
\sigma \rho'   + \bigg(\rmi \rmk_R\rho + \pdv{\rho}{R} \bigg) \rmv_\rmR'  + \pdv{\rho}{Z} \rmv_\rmz'  + \rho\pdv{\rmv_\rmz'}{Z}  = 0,
\end{equation} 

\begin{align} 
\sigma \Pi'   + \bigg(  \rmi \rmk_R \Pi\gamma_\rmB  + \pdv{\Pi}{R} &- \rho\pdv{\cs^2}{R}\bigg) \rmv_\rmR'  + \pdv{\Pi}{Z} \rmv_\rmz'  \nn \\
& + \gamma_\rmB \Pi\pdv{\rm\rmv_\rmz'}{Z} +  \frac{\rmB_\phi}{\mu_0}\eta \nabla^2 \rmB_\phi' = 0.
\label{eq:l1_}
\end{align} 
Note that the magnetic tension force vanishes for a purely toroidal field and neglecting curvature terms. The perturbed azimuthal magnetic field $\rmB_\phi'$ can be expressed as
\begin{equation}
\rmB_\phi'= \frac{\mu_0}{\rmB_\phi}(\Pi'-\rho'\cs^2). 
\end{equation} 
When $\beta \to \infty$ and hence $\Pi' \to P'$, the set of linearized equations recover the hydrodynamic limit \citep{ly15}. For finite magnetic field strengths and $\eta=0$, these equations describe the ideal MHD limit. For $\eta > 0$, the set of equations describe non-ideal MHD.

\subsection{Analytical Solutions} \label{sec:analy}
Analytic solutions can be obtained in the limit of ideal MHD ($\eta=0$) with a large and constant $\beta$, by solving the  linearized equations with polynomial solutions. To do so, we make the following simplifying assumptions \citep{ly15}: 
\begin{enumerate}

\item A fully radially local approximation ($\partial /\partial R =0$) for background disks. However, the vertical shear that originates from the radial temperature profile ($\partial T /\partial R$) is retained. 
\\
\item We set $\Omega = \Omega_{\rm K}$ and $\kappa = \Omega_{\rm K}$ where they appear explicitly and without vertical derivatives. As seen in Equation \eqref{eq:omg}, the Keplerian approximation is only valid for large $\beta$. A strong magnetic field will lead to substantial deviation from $\Omega_{\rm K}$. 
\\
\item In the hydrodynamic limit, the VSI is an overstability due to the destabilization of inertial waves \citep{bl15}. We expect a similar result for weak magnetizations and consider low frequency modes with $|\sigma|\ll \Omega_{\rm K}$ to filter out acoustic waves \citep{lp93}. 
\\
\item We consider thin disks and set
\begin{alignat}{2}
& \frac{1}{\rho}\frac{\p \Pi}{\p Z} &&=- \OmK^2 Z, \nonumber \\
& \frac{\p \ln{\rho}}{\p Z} &&=  -\frac{\beta}{1+\beta}\frac{Z}{H^2}, \nonumber \\ 
& R\frac{\p \Omega^2}{\p Z} &&= - \qT \frac{Z}{R}\OmK^2. 
\label{eq:eqm_approx}
\end{alignat}
\end{enumerate}

To non-dimensionalize the perturbation equations, we choose the appropriate scalings for timescales, lengthscales, and velocities to be the Keplerian orbital time $\Omega_{\rm K}^{-1}$, background sound speed $\cs$, and pressure scale-height $H$. We thus write
\begin{equation}
\sigma \rightarrow  \Omega_{\rm K} \sigma^*, \quad Z \rightarrow H Z^*, \quad k_R \rightarrow K/H, 
\label{eq:ast}
\end{equation}
where scaled variables are denoted by an asterisk and are omitted below. 
Equations \eqref{eq:l1} can be combined into a single second-order ordinary differential equation, 
\begin{equation}
\dv[2]{\rmv_\rmz'}{Z} - a_1 \dv{\rmv_\rmz'}{Z}Z + \rmv_\rmz'(a_2  - a_3 Z^2) = 0,
\label{eq:ode1}
\end{equation}
where $a_1$, $a_2$ and $a_3$ are constants in $Z$ and are defined by 
\begin{align}
a_1  &= \frac{\beta}{1+\beta} - \mathrm{i}K q_T h,  \nonumber \\
a_2  &= -\frac{\beta}{2+\beta} - \sigma^2 K^2 + \mathrm{i}K q_T h, \nonumber \\
a_3  &= \frac{\beta}{2+\beta} (\gamma_\rmB -1)\bigg(K^2 + \mathrm{i}K q_T h\bigg).
\label{eq:coeff}
\end{align}
Equation \eqref{eq:ode1} is equivalent to the hydrodynamic result derived by \citet[see their Equation (29)]{ly15} without cooling and assuming $K\gg 1$. This can be seen by setting $Z \equiv  \tilde{Z}\sqrt{(1+\beta)/\beta}$, $K \equiv \tilde{K}\sqrt{\beta/(1+\beta)}$, and $h \equiv \tilde{h}\sqrt{\beta/(1+\beta)}$. However, they also show that in the hydrodynamic limit, neglecting the radial structure of disk while retaining vertical shear is only valid for gas that is nearly isothermal in its thermodynamic response. We thus expect Equation \eqref{eq:ode1} to only apply for weak magnetizations where $\gamma_\rmB \to 1$. In the limit $\beta\to \infty$ and hence $a_3\to 0$, we recover the equation for the VSI in locally isothermal disks \citep{ly15,bl15}. When there is no vertical shear ($q_T=0$), Equation \eqref{eq:ode1} is analogous to that obtained by \cite{lp93} for axisymmetric waves in adiabatic disks.     

We bring Equation \eqref{eq:ode1} into a form that can be solved analytically \citep{lp93}. Define a variable $Y(Z)$ 
\begin{equation} 
Y(Z) = \rmv_\rmz'(Z)\exp(-Z^2\xi/2),
\label{eq:yz}
\end{equation} 
where $\xi$ is a constant to be determined. Plugging this into Equation \eqref{eq:ode1} yields an ordinary differential equation in $Y$,
\begin{equation}\label{eq:ode_temp}
\dv[2]{Y}{Z} - (a_1-2\xi)Z\dv{Y}{Z}  + Y[(a_2+\xi) + (\xi^2-a_1\xi-a_3)Z^2] =0.
\end{equation}
Then choose $\xi$ to eliminate the $Z^2$ term, 
\begin{equation}\label{xi_def}
\xi = \frac{1}{2}a_1 \pm \frac{1}{2}\sqrt{a_1^2+4a_3},
\end{equation}
which brings Equation \eqref{eq:ode_temp} into
\begin{equation}
\dv[2]{Y}{Z} - (a_1-2\xi)Z\dv{Y}{Z}  + Y(a_2+\xi) = 0.
\label{eq:ode}
\end{equation}
We proceed to represent solutions of $Y$ by $Y_n(Z)$, an $n$th order polynomial in $Z$,
\begin{equation}
Y_n(Z) = \sum_{m=0}^n B_m Z^m,
\label{eq:y}
\end{equation}
where $B_m$ are constant coefficients. Substituting this into Equation \eqref{eq:ode}, we arrive at the recurrence relation between $B_{m}$ and $\rmB_{m+2}$,
\begin{equation}
B_{m+2}= \frac{(a_1-2\xi_n)m-(a_2+\xi_n)}{(m+1)(m+2)}B_m,
\label{eq:rec}
\end{equation} 
where we have relabled $\xi \to \xi_n$. 
For physical solutions to Equation \eqref{eq:ode}, 
we demand the vertical kinetic energy density of perturbations to remain bound, i.e. $\rho |\rmv_\rmz'|^2$ approaches zero for large $|Z|$. Then $Y_n(Z)$ should be a polynomial, as assumed. Thus $B_{n+2}=0$ when $m=n$, or
\begin{equation}
(a_1-2\xi_n)n = a_2+\xi_n. 
\label{eq:dr}
\end{equation} 
This relation can be also quickly obtained by recognizing Equation \eqref{eq:ode} as the Hermite differential equation \citep{bl15}. We choose the negative root in Equation \eqref{xi_def} for physical solutions, giving 
\begin{equation}
\xi_n = \frac{1}{2}a_1 - \frac{1}{2}\sqrt{a_1^2+4a_3}.
\label{xi_n_sol}
\end{equation} 
For a given set of disk parameters $(q,\beta, h)$, Equations \eqref{eq:coeff}, \eqref{eq:dr}, and \eqref{xi_n_sol} can be readily solved to obtain the dispersion relation for the complex frequency $\sigma_n = \sigma_n(K; n)$ of the $n\mathrm{th}$ mode. The corresponding eigenfunction $\rmv_\rmz^\prime$ is then given by Equations \eqref{eq:yz}, \eqref{eq:y}, and \eqref{eq:rec}. We note that the vertical shear rate increases without bound with height in thin disks (Equation \eqref{eq:eqm_approx}), leading to unbound growth rates (as seen in Figure \ref{fig:N13} discussed below), which violates the low frequency approximation.

\subsection{Numerical Solutions}\label{sec:num}
We also solve the full linearized Equations \eqref{eq:l1} -- \eqref{eq:l1_} numerically. The equations can be written in a form of standard matrix eigenvalue problem,
\begin{equation}
 \mathcal{L}  \cdot  \mathbf{X}  + \sigma \cdot \mathbf{X}  = 0,
\end{equation} 
where $\sigma$ is the eigenvalue,  $\mathcal{L}$ is a $5 \times 5$ matrix of linear operators, and $\mathbf{X} =[\rmv'_\rmR, \rmv'_\phi,\rmv'_\rmz, \rho', \Pi' ]^\mathrm{T}$ is a vector of eigenfunctions. We use \textsc{dedalus}\footnote{\url{https://dedalus-project.org/}} \citep{burns_etal20}, a general purpose spectral code for differential equations, to solve the linear eigenvalue problem. We employ a Chebyshev collocation grid of $N=150$ points. Spurious solutions are filtered out and numerical convergence is verified against double-resolution calculations with $N=300$ by using the \textsc{eigentools }  package\footnote{\url{https://github.com/DedalusProject/eigentools}}.

Unlike the analytic model in \S \ref{sec:analy}, where the disk surface extends to infinity due to the vertically isothermal background state, for numerical solutions we consider a finite vertical domain with $Z\in[-5, 5]H$. The boundary conditions imposed at the disk surfaces are 
\begin{equation}
\rmv_\rmz' = 0, \quad \pdv{\Pi'}{Z} =0.
\end{equation} 
The former condition is adopted in the limit of ideal MHD, and the latter being additional conditions when resistivity is included. We have experimented with boundary conditions and found that our main findings are insensitive to them. 

\section{Results}\label{sec:re}

\begin{figure}
\centering
\includegraphics[width=0.45\textwidth]{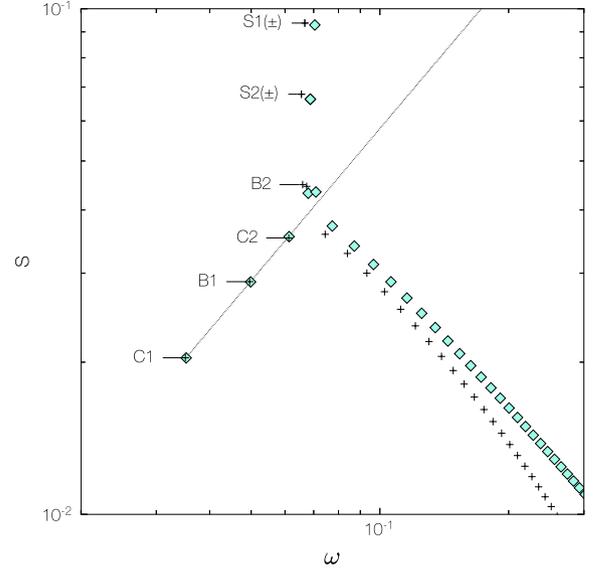}
\caption{Comparison of growth rates $s$ and oscillation frequencies $\omega$ of N13 and this work ($\beta_\phi \rightarrow \infty$). Line denotes analytical solutions obtained in \S\ref{sec:analy}. Diamonds denote numerical solutions obtained in N13. Crosses denote numerical solutions obtained in \S\ref{sec:num}. 
Labels B, C, S represent breathing, corrugation, and surface modes, respectively. Numbers represent fundamental and first overtone modes. Modes resides in the lower-right are high-order body modes.}
\label{fig:N13}
\end{figure}

In this section, we present example solutions in the hydrodynamic limit (\S\ref{sec:hydro}), the ideal MHD limit (\S\ref{sec:ideal}), and the non-ideal MHD limit (\S\ref{sec:nonideal}). The fiducial parameter values are $h = 0.05$, $\qT=1$, $\qD=1.5$, and $K=35$. Note that $\qT=1$ gives a constant disk aspect ratio $h$. We present the non-dimensionlized perturbed quantities as in Equation \eqref{eq:ast}. Plasma beta parameters associated with azimuthal and vertical fields are denoted by $\beta_{\phi} =2\mu_0 P/B_{\phi}^2$ and $\beta_Z =2\mu_0 P/B_Z^2$. We denote midplane plasma beta parameters by $\beta_{\phi0}$ and $\beta_{Z0}$, and midplane Els\"{a}sser number $\Lambda_0$. For reference, values of $\beta_{Z} \sim 10^4$ and $\beta_{\phi} \sim 10^2$ are found to account for the accretion rate in PPDs \citep{simon_etal13a,bai15}. Note, however, as in the discussion above in this section we only consider azimuthal fields, so that $\beta=\beta_\phi$. Poloidal fields will be explored in Section \ref{sec:po}.

We follow N13 to denote breathing, corrugation, and surface modes as B, C, and S, respectively, with numbers 1 and 2 representing the fundamental and first overtone modes. For clarity, analytic solutions which have discrete modes are plotted as continuous curves. 

\begin{figure*}
\includegraphics[width=0.95\textwidth]{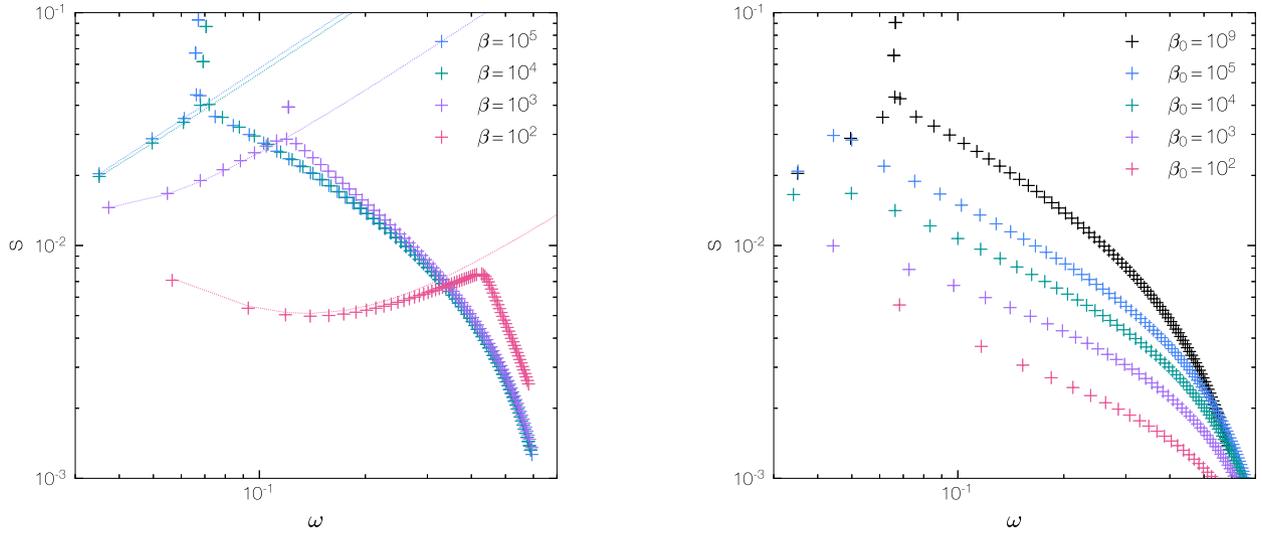}
\caption{Growth rates $s$ and oscillation frequencies $\omega$ of unstable modes at discrete plasma $\beta_\phi$ or $\beta_{\phi0}$. Left: a constant-$\beta$ disk. Right: a constant-$\rmB_\phi$ disk. Curves denote analytic solutions (\S\ref{sec:analy}) and crosses denote numerical solutions (\S\ref{sec:num}). }
\label{fig:ideal_beta}
\end{figure*}

\begin{figure*}
\includegraphics[width=1\textwidth]{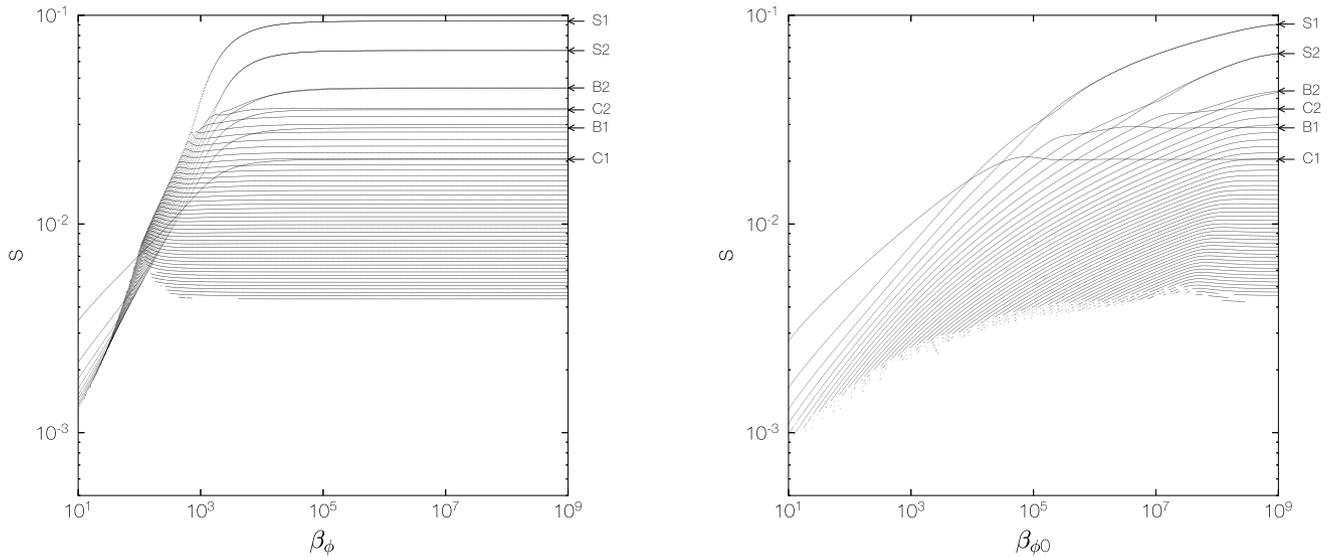}
\caption{Growth rates $s$ of all unstable modes as a function of plasma $\beta_\phi$ or $\beta_{\phi0}$. Left: a constant-$\beta$ disk. Right: a constant-$\rmB_\phi$ disk.}
\label{fig:ideal_mode}
\end{figure*}

\subsection{The Hydrodynamic Limit} \label{sec:hydro}

\begin{figure*}
\includegraphics[width=0.95\textwidth]{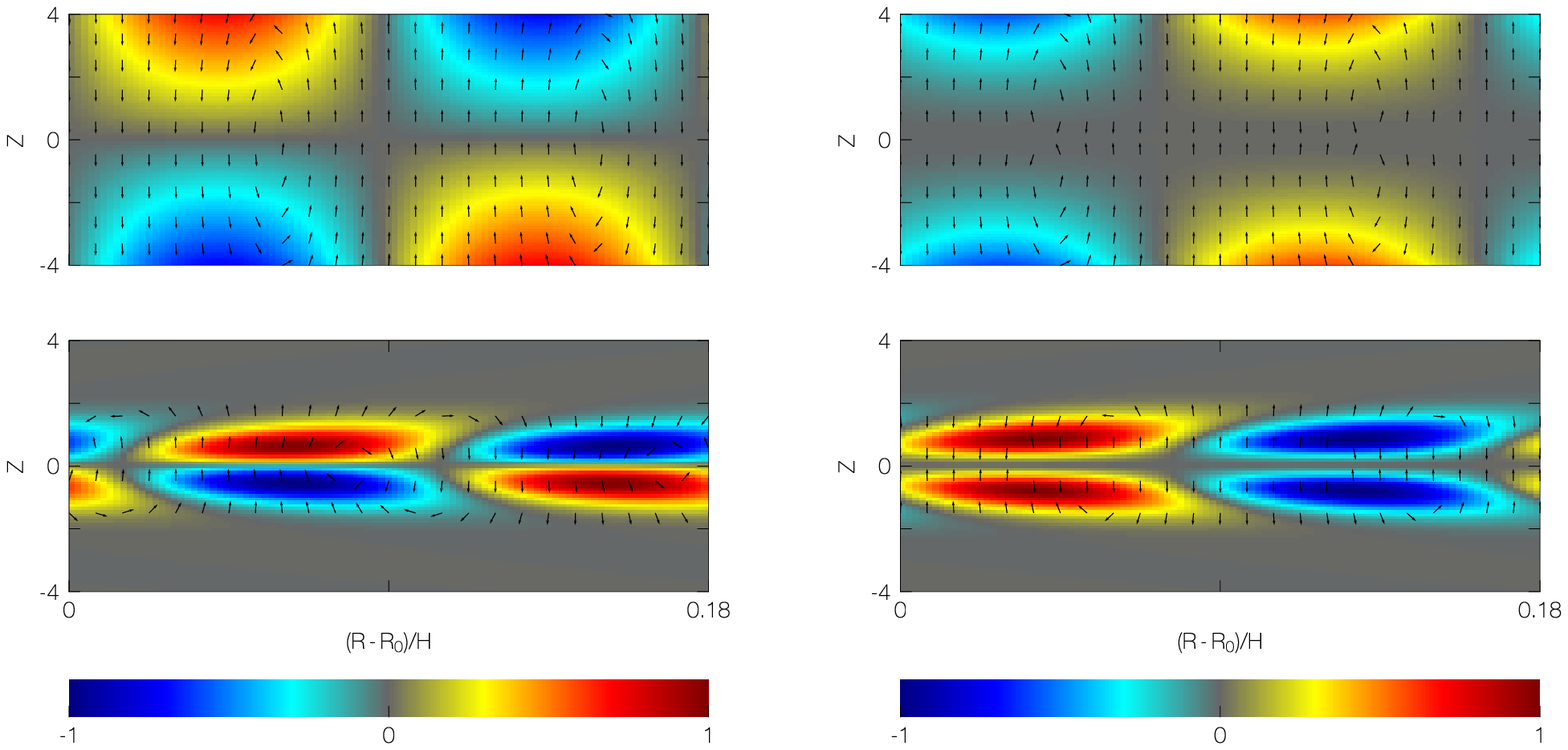}
\caption{Fundamental corrugation modes C1 (left) and fundamental breathing modes B1 (right) at $\beta_\phi=10^5$ (top) and $\beta_\phi=10^2$ (bottom) in constant-$\beta$ disks. Contours show the magnetic field perturbations, $\Re \{\rmB_\phi'\exp[\rmi K(R-R_0)]\}/B_\phi$, normalized by its maximum. Arrows denote perturbed velocity vectors ($\rmv_R', \rmv_Z'$). The radial interval of $K(R-R_0)=2\pi$ corresponds to $0.18H$.} 
\label{fig:igenf}
\end{figure*}

We first compare our numerical and analytical solutions with N13, who considered purely hydrodynamic disks. To this end, we compute the numerical solutions to Equations \eqref{eq:l1} -- \eqref{eq:l1_} and analytical solutions given via Equation \eqref{eq:dr} in the hydrodynamic limit ($\beta_\phi\to\infty$), and compare with numerical solutions to Equation (39) in N13. Note that N13 employed the anelastic approximation $(\p \rho/\p t=0)$, while we account for full compressibility in numerical solutions. The results are shown in Figure \ref{fig:N13}.

For the fundamental and first overtone breathing and corrugation modes (B1, B2, C1, and C2), all three methods yield consistent results. For surface modes and higher-order body modes in the lower right of Figure \ref{fig:N13}, the two numerical solutions also show consistency, especially at low oscillation frequencies. The analytic solutions for these higher-order body modes do not match with numerical solutions due to the lack of a disk surface in the former \citep{bl15}. In practice, the growth rate is limited by the maximum vertical shear rate within the domain \citep{ly15}, as reflected in the numerical solutions. Overall, the comparison is satisfactory. 

\subsection{The Ideal MHD Limit}\label{sec:ideal}

In the ideal MHD limit, we examine the behaviour of the VSI modes as functions of disk magnetizations $\beta_\phi$ (\S\ref{sec:dm}), radial wavenumbers $K$ (\S\ref{sec:WN}), and disk aspect ratios $h$ (\S\ref{sec:h}).

\subsubsection{Disk magnetization}\label{sec:dm}

We now examine the strengths of toroidal magnetic fields on the VSI. In Figure \ref{fig:ideal_beta}, we show example growth rates and frequencies for constant-$\beta$ disks in the left panel and constant-$\rmB_\phi$ disks in the right panel. Similarly, Figure \ref{fig:ideal_mode} shows how growth rates of various modes vary with plasma beta. The curves without labels are high-order body modes. 

We highlight three major findings. Firstly, strong magnetization reduces the VSI growth. Physically, this is because the gas and the magnetic fields are perfectly coupled in the limit of ideal MHD, so that magnetic fields impede the free movement of the perturbed gas. Furthermore, surface modes are the first to vanish with strong magnetization. This can be understood by the fact that the  stabilizing vertical buoyancy scales as $Z^2$ for both models as seen in Equations \eqref{eq:Nzbeta} and \eqref{eq:Nzbphi}, hence the gas is subject to stronger stabilization at the disk surface. Finally, the critical $\beta_\phi$ to recover hydrodynamic results for a constant-$\beta$ disk, $\beta_\phi \gtrsim 10^5$, is smaller than that for the midplane value in a constant-$\rmB_\phi$ disk, $\beta_{\phi0} \gtrsim10^9$. This is because in a constant-$\rmB_\phi$ disk, Equation \eqref{eq:beta} shows that $\beta_\phi$ decreases with height, so the vertically averaged $\beta$ is smaller than its midplane value. 

Figure \ref{fig:igenf} shows the flow structure in a constant-$\beta$ disk. The radial domain of $K(R-R_0)=2\pi$ corresponds to an interval of $0.18H$. The left panels show the fundamental corrugation modes in disks with $\beta_\phi=10^5$ (top) and $\beta_\phi=10^2$ (bottom). The perturbed vertical velocities show even symmetry about the midplane. The right panels are corresponding fundamental breathing modes, where the perturbed vertical velocities have odd symmetry. The contours show the magnetic field perturbations $\Re \{\rmB_\phi'\exp[\rmi K(R-R_0)]\}/B_\phi$ and is normalized by its maximum value. The perturbed magnetic fields possess opposite symmetry to perturbed vertical velocities. Importantly, we find that strong magnetization confines VSI activity towards the midplane since the stabilizing vertical buoyancy increases with height. The same arguments also apply to a constant-$\rmB_\phi$ disk.
 
\subsubsection{Radial wavenumber}\label{sec:WN}
\begin{figure*}
\includegraphics[width=0.9\textwidth]{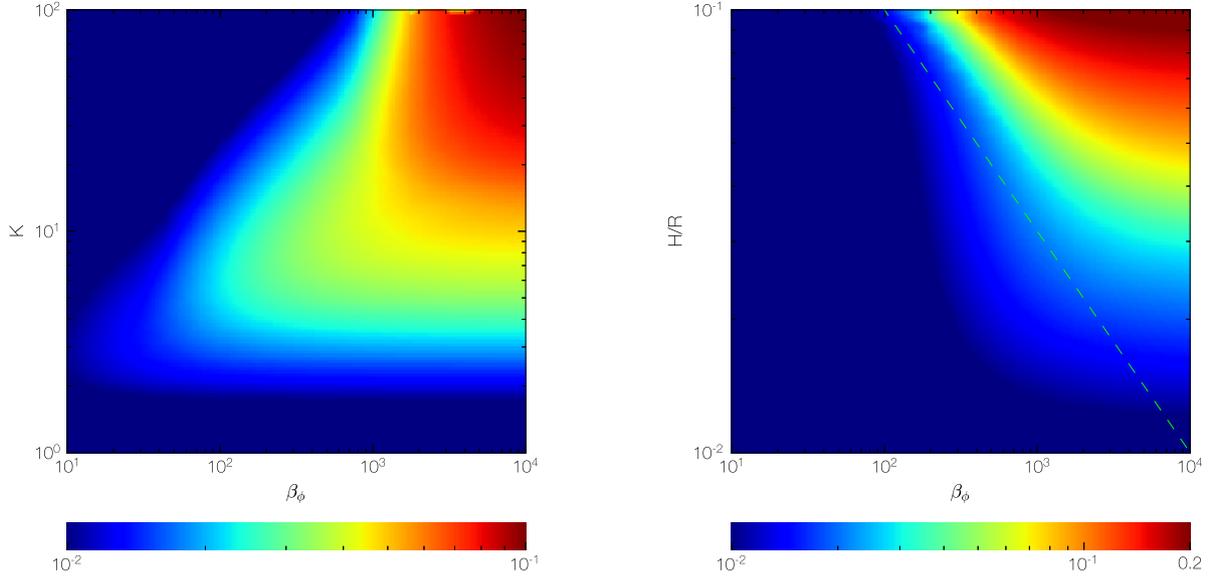}
\caption{Contours of maximum growth rates in logarithmic scale as functions of $\beta_\phi$ and radial wave number $K$ (left) or disk aspect ratio $h$ (right) in a constant-$\beta$ disk. Dashed line represents $\beta_\phi=h^{-2}$.}
\label{fig:ideal_c}
\end{figure*}

The left panel of Figure \ref{fig:ideal_c} depicts contours of maximum growth rates as a function of $\beta_\phi$ and radial wavenumber $K$ for constant-$\beta$ disks.  
A critical $\beta_\textrm{c} \sim 10^3$ can be defined to separate the disks into two regimes. In constant-$\rmB_\phi$ disks it is $\beta_\textrm{c} \sim 10^5$. For $\beta_\phi \gtrsim \beta_\textrm{c}$, the maximum growth rate is a monotonically increasing function of $K$, whereas for $\beta_\phi \lesssim \beta_\textrm{c}$, the maximum growth rate peaks at some intermediate $K$. We explain below that $\beta_\textrm{c}$ is in fact the critical disk magnetization below which surface modes are quenched. In Figure \ref{fig:ideal_mode}, we see that at $\beta_\phi \gtrsim \beta_\textrm{c}$, surface modes, which prefer very small radial wavelengths, dominate the maximum growth rates resulting in fast growth rates at large $K$. At $\beta_\phi \lesssim \beta_\textrm{c}$, surface modes are suppressed, while body modes that prefer longer radial wavelengths persist, and thus maximum growth rates appear at intermediate radial wavenumbers. 

\subsubsection{Disk aspect ratio}\label{sec:h} 

In the right panel of Figure \ref{fig:ideal_c}, we show contours of maximum growth rates as functions of $\beta_\phi$ and disk aspect ratio $h$, again for constant-$\beta$ disks. The maximum growth rates increase with the disk aspect ratio for a given $\beta_\phi$.  
Requiring modes to fit into the vertical height of the disk, a lower limit can be placed on $\beta_\phi$ for the VSI to operate, $\beta_\textrm{min} \gtrsim (-R\p \ln \Omega/\p Z)^{-2}$ (LP18). This is set by the vertical shear rate, and can be simplified to $\beta_\textrm{min} \gtrsim h^{-2}$ using Equation \eqref{eq:eqm_approx}. Note, however, this criterion was derived for purely poloidal background fields in a local approximation, while we consider purely toroidal magnetic fields in a vertically global disk. Nevertheless, we find the local condition $\beta_\phi = h^{-2}$, shown in Figure \ref{fig:ideal_c} as the dashed line, successfully predicts the quenching of the VSI in our disk model. 

\subsection{The Non-ideal MHD Limit} \label{sec:nonideal}

\begin{figure*}
\includegraphics[width=0.9\textwidth]{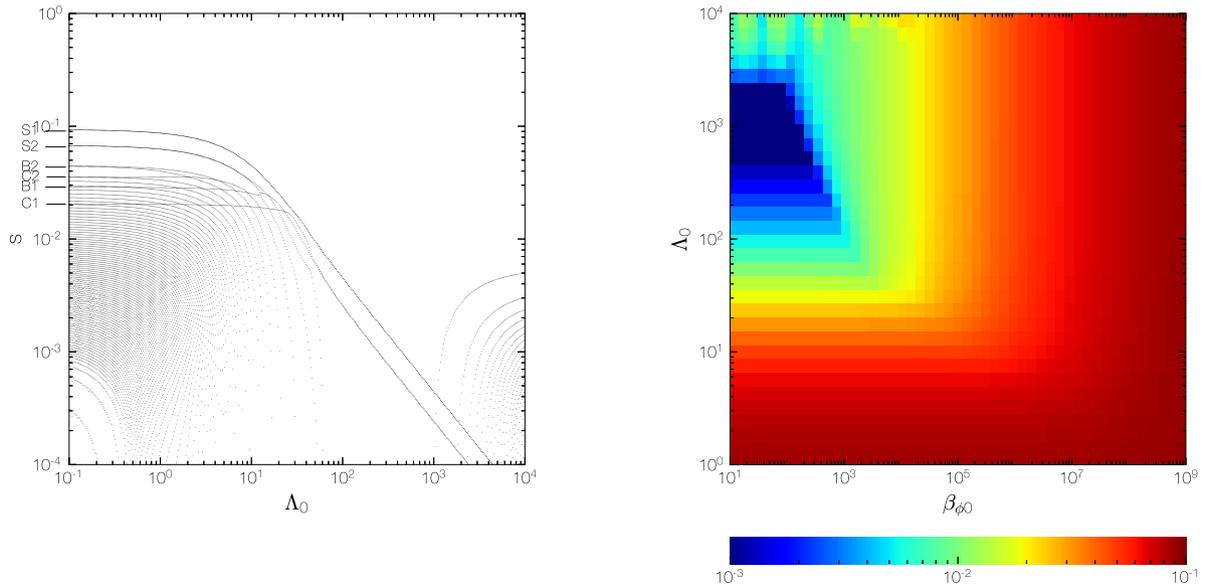}
\caption{The effect of Ohmic resistivity on the VSI growth rate. Left: growth rate $s$ versus Ohmic Els\"{a}sser number $\Lambda_0$ at $\beta_{\phi0}=10^2$. Right: Contour of maximum growth rates in logarithmic scales as functions of plasma $\beta_{\phi0}$ and Ohmic Els\"{a}sser number $\Lambda_0$.}
\label{fig:ohmic}
\end{figure*}

In this subsection, we show that the VSI can be revived when non-ideal MHD effects are included. To assure the existence of equilibrium solutions, a constant-$\rmB_\phi$ disk model is employed (\S\ref{sec:cBphi}). With purely toroidal magnetic fields in axisymmetric disks, the three non-ideal MHD effects reduce to only Ohmic resistivity, because the Hall effect vanishes, and ambipolar diffusion acts as an effective resistivity with field dependency (\S\ref{sec:toraxis}). Therefore, we only explore the dependency of Ohmic Els\"{a}sser number $\Lambda$, while we expect the same results apply to ambipolar diffusion. We take the diffusivity $\eta$ to be constant so that the Els\"{a}sser number increases with $|Z|$, as shown in Equation \eqref{eq:elsasser}. 

In the left panel of Figure \ref{fig:ohmic}, we show the growth rates of all unstable modes as a function $\Lambda_0$ at $\beta_{\phi0}=10^2$. The labels correspond to surface and body modes in the hydrodynamic limit (Figure \ref{fig:N13}), which is recovered for small $\Lambda_0$ or strong Ohmic resistivity. On the other hand, $\Lambda_0\to\infty$ tends to the ideal MHD limit. We find for $\Lambda_0$ $\gtrsim 10^3$, the surface modes vanish and the growth rate of body modes is significantly reduced due to strong magnetization.
As $\Lambda_0$ declines from larger  values, the growth rates of these body modes drop at around $\Lambda_0 \sim 10^3$, then they re-emerge and converge to hydrodynamic results. 
The growth rates of all modes converge to hydrodynamic results for $\Lambda_0 \lesssim 10$, with the transition starting at $\Lambda_0 \sim 10^2$. Local analyses demonstrate that the  stabilizing effect by magnetic fields will be overcome by  magnetic diffusion when $\Lambda \lesssim h^{-1}$ ($=20$ in our fiducial disk) for a mode with growth rate $\sim h \Omega$ (LP18)\footnote{Equation (64) of LP18 contains a typographical error, the corrected expression is $E_\eta \la 1/q$ (H. Latter, private communication).}. This is in agreement with our results, though the growth rates from our solutions are only reduced rather than completely suppressed.

The right panel of Figure \ref{fig:ohmic} shows the maximum growth rates as functions of $\beta_{\phi0}$ and $\Lambda_0$. For $\beta_{\phi0}>10^3$, the maximum growth rate is a monotonically decreasing function with increasing $\Lambda_0$, whereas for $\beta_{\phi0}<10^3$, the maximum growth rate has its minimum resides at some intermediate $\Lambda_0$, corresponding to the left panel of Figure \ref{fig:ohmic}. The hydrodynamic result is recovered for sufficiently weak fields ($\beta_{\phi0}\gtrsim 10^5$) \emph{or} sufficiently strong resistivity ($\lambda_0\lesssim 10$). 

\section{Purely Poloidal Background Magnetic Fields}\label{sec:po}

The above analyses focus on disks threaded by a toroidal magnetic field, which is expected to dominate over poloidal field strengths in PPDs \citep[e.g.][]{bai17, bethune_etal17,cb20}. However, the presence of a poloidal field, even weak, can lead to new effects such as MHD disk winds and the MRI \citep{bai13,simon_etal13b,gressel_etal20}. Specifically, the surface layers in outer regions of PPDs are likely sufficiently ionized by stellar FUV radiation to trigger the MRI \citep{pc11, simon_etal13a, simon_etal13b,bai15}. These regions are also prone to the VSI since the vertical shear rate increases with height. In this section, we investigate the VSI modes in a disk with purely poloidal magnetic fields, $\bB=(B_R,0,B_Z)$. In \S\ref{sec:gb}, we study the effects of poloidal magnetic fields on the VSI in a vertically global disk model. In \S\ref{sec:MRI}, we compare the MRI with the VSI in a local disk model. Although toroidal fields are absent in the background, it is allowed in the perturbed state. 

\subsection{Vertically Global Model}\label{sec:gb}

We make several simplifying assumptions to establish disk equilibria with a purely poloidal magnetic field:   
\begin{enumerate}
\item The Lorentz force in the momentum equation is ignored because the disk is weakly magnetized. We therefore use  equilibrium solutions for $\Omega$ and $\rho$ in the hydrodynamic limit (Equations \eqref{rhoRz}--\eqref{eq:omg} with $\beta \to \infty$, see also N13).
\item We assume thin disks and consider $h \ll 1$. 
\item A constant background vertical magnetic field $\rmB_\rmz$ is assumed, and we seek the required equilibrium radial magnetic field $\rmB_\rmR$, as follows. 
\end{enumerate} 

The equilibrium magnetic fields must satisfy solenoidal condition and induction equation. Considering only Ohmic resistivity with a constant diffusivity, the equilibrium induction equation is 
\begin{equation}
0 = (\bB \cdot \nabla) \bv_\phi - (\bv_\phi\cdot\nabla)\bB+ \eta \nabla^2 \bB,
\label{eq:bp_idc}
\end{equation}
where $\bm{v}_\phi = R\Omega \hat{\bm{\phi}}$. 
In the thin disk approximation, the gradients of $\rmv_\phi$ are 
\begin{equation}
\pdv{\rmv_\phi}{Z} \simeq - \frac{1}{2}\OmK \qT\frac{Z}{R},
\end{equation} 
\begin{equation}
\pdv{\rmv_\phi}{R} \simeq -\frac{1}{2}\OmK.
\end{equation} 
The induction equation can be satisfied by the radial field,
\begin{equation}
\rmB_\rmR= -\frac{1}{3}\qT\frac{Z}{R}\rmB_\rmz. 
\label{eq:BR}
\end{equation}
Notice that this field configuration is not subject to Ohmic diffusion and satisfies the solenoidal condition, 
\begin{equation}
\frac{1}{R}\pdv{(R\rmB_\rmR)}{R} + \pdv{\rmB_\rmz}{Z}=0.
\label{eq:bp_sol}
\end{equation} 
Therefore, an approximate equilibrium magnetic field configuration is obtained. The equilibrium solution, Equation \eqref{eq:BR}, resembles Equation (50) in LP18. Since $|\rmB_\rmR|\sim O(h) \rm|B_\rmz|$, the strength of the magnetic field is dominated by the vertical field, which is assumed a constant so that this is similar to the constant-$\rmB_\phi$ disks considered in \S\ref{sec:eqmdisk}. 

\begin{figure*}
\centering
\includegraphics[width=1\textwidth]{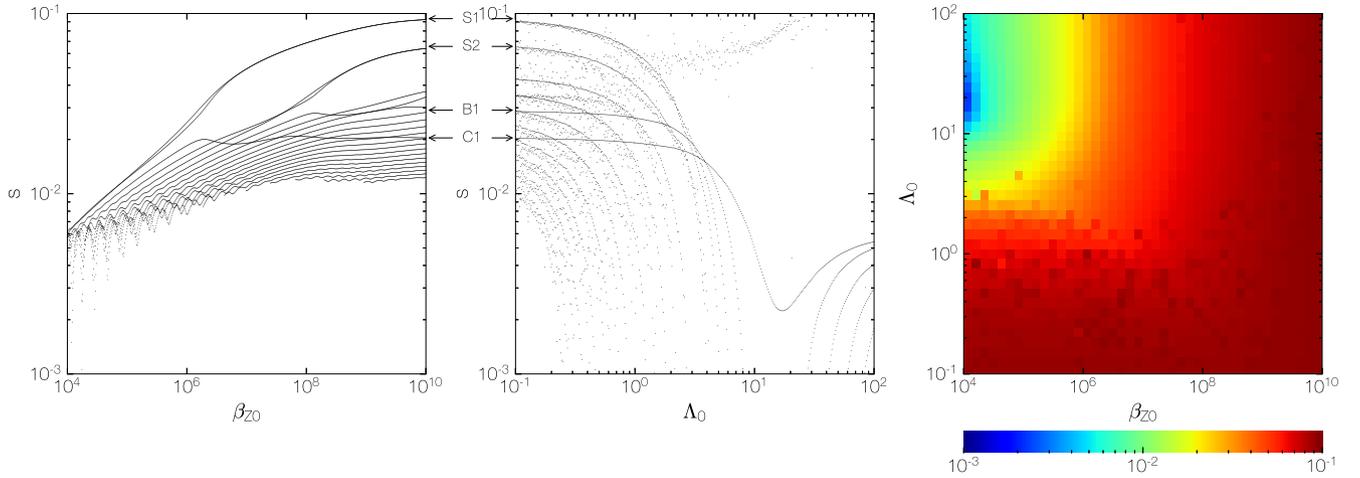}
\caption{Left: growth rates $s$ of all unstable modes as a function of midplane $\beta_{Z0}$ in the ideal MHD limit. Middle: growth rates $s$ versus midplane Ohmic Els\"{a}sser number $\Lambda_0$ at $\beta_{Z0}=10^4$. Right: Contour of maximum growth rates as functions of $\beta_{Z0}$ and Ohmic Els\"{a}sser number $\Lambda_0$.}
\label{fig:p}
\end{figure*}

With a poloidal field it is not possible to map the problem to adiabatic hydrodynamics. We therefore work with the MHD equations directly. The radial derivatives of background quantities are omitted. The set of linearized equations are 
\begin{equation}
\sigma \frac{\rho'}{\rho} + \ik_\rmR  \rmv_\rmR' + \dv{\rmv_\rmz'}{Z} + \dv{\ln \rho}{Z}\rmv_\rmz'=0,
\label{eq:l2}
\end{equation} 
\begin{align}
\sigma \rmv_\rmR' -2\Omega\rmv_\phi' &+ \ik_\rmR\frac{P'}{\rho} +  \frac{\ik_\rmR}{\mu_0\rho} \bB \cdot \bB' \nn \\
& - \frac{1}{\mu_0\rho} \bigg[\ik_\rmR \rmB_\rmR \rmB_\rmR' + \rmB_\rmz \dv{\rmB_\rmR'}{Z}+   \dv{\rmB_\rmR}{\rmz} \rmB_\rmz'  \bigg] =0,
\end{align} 
\begin{align}
\sigma \rmv_\phi' + \frac{\kappa^2}{2\Omega}\rmv_\rmR' &+ \dv{\rmv_\phi}{Z}\rmv_\rmz'  \nn \\
&- \frac{1}{\mu_0\rho}   \bigg[\ik_\rmR \rmB_\rmR \rmB_\phi' + \rmB_\rmz \dv{\rmB_\phi'}{Z} \bigg]=0,
\end{align} 
\begin{align}
\sigma\rmv_\rmz' +\frac{1}{\rho}\dv{P'}{Z} &-\frac{\rho'}{\rho^2} \dv{P}{Z}  +  \frac{1}{\mu_0\rho} \pdv{\bB \cdot \bB}{Z}  '  \nn \\
& - \frac{1}{\mu_0\rho}   \bigg[\ik_\rmR \rmB_\rmR \rmB_\rmz' + \rmB_\rmz \dv{\rmB_\rmz'}{Z}\bigg]=0,
\end{align} 
\begin{align}
\sigma\rmB_\rmR' &- \bigg[\ik_\rmR \rmB_\rmR + \rmB_\rmz \dv{}{\rmz} \bigg]\rmv_\rmR'  \nn \\
&+ \bigg[\ik_\rmR \rmv_\rmR' + \dv{\rmv_\rmz'}{\rmz} + \rmv_\rmz'\pdv{}{\rmz} \bigg]  \rmB_\rmR + \bigg[\rmk_\rmR^2  - \dv[2]{}{Z} \bigg] \eta \rmB_\rmR' =0,
\end{align} 
\begin{align}
\sigma \rmB_\phi' &- \bigg[\ik_\rmR \rmB_\rmR + \rmB_\rmz \dv{}{\rmz} \bigg]\rmv_\phi'  \nn \\
&-\bigg[\frac{\kappa^2}{2\Omega}-2\Omega\bigg] \rmB_\rmR'  - \pdv{\rmv_\phi}{Z} \rmB_\rmz' + \bigg[\rmk_\rmR^2  - \dv[2]{}{Z} \bigg]\eta\rmB_\phi' =0,
\end{align} 
\begin{align}
\sigma\rmB_\rmz' &- \bigg[\ik_\rmR \rmB_\rmR + \rmB_\rmz \dv{}{\rmz} \bigg]\rmv_\rmz'  \nn \\
&+  \bigg[\ik_\rmR \rmv_\rmR' + \dv{\rmv_\rmz'}{\rmz}\bigg]  \rmB_\rmz + \bigg[\rmk_\rmR^2  - \dv[2]{}{Z} \bigg] \eta \rmB_\rmz' =0,
\end{align} 
\begin{equation}
P' = \rho' \cs^2.
\label{eq:l2_}
\end{equation}
Note that there is now a magnetic tension force in the momentum equations. 

We solve the linear eigenvalue problem, in its dimensionless form, numerically using the spectral method described in \S\ref{sec:num}. A resolution of $N=100$ is used. Spurious solutions are filtered out with double resolution calculations. Boundary conditions imposed at upper and lower disk surfaces for ideal MHD limit are (\citealp{gammie94,sano99})
\begin{equation}
\rho'=0, \quad \rmB'_{\rmR} =0, \quad \rmB'_{\phi}  = 0,
\end{equation}
Additional conditions are imposed with Ohmic resistivity, 
\begin{equation}
P^{'} = 0, \quad \rmB'_{\rmz} = 0.
\end{equation}

The quantities $\sigma$, $Z$, $K$ reported below are non-dimensionlized as in Equation \eqref{eq:ast}. With the inclusion of vertical magnetic fields, MRI modes emerge in the numerical solutions. Unlike the VSI, however, MRI modes are not overstable even with resistivity, i.e. $\Im(\sigma) \equiv \omega = 0$ \citep{sano99}; while it can be seen 
in Figure \ref{fig:ideal_beta} that VSI modes generally have $0.01< \omega < 1$, hence the MRI modes do not contaminate the numerical results of the VSI.

The left panel of Figure \ref{fig:p} shows growth rates of all unstable modes as a function of disk magnetization $\beta_{Z0}$ in the ideal MHD limit. Consistent with purely the toroidal field model, strong magnetization suppresses VSI growth and surface modes are the first to vanish with increasing field strengths. 
The critical $\beta_{Z0}$ to recover hydrodynamic results is even larger, $\beta_{Z0} \gtrsim 10^{10}$, because the total magnetic field strength increases over height as radial magnetic field develops away from the midplane in Equation \eqref{eq:BR}. 

Growth rates including Ohmic resistivity are shown in the middle and right panels of Figure \ref{fig:p}. The middle panel shows the growth rates of all unstable modes as a function of Ohmic Els\"{a}sser number $\Lambda_0$ at $\beta_{Z0}=10^4$. Nearly all the VSI modes are suppressed when Ohmic resistivity is weak at  $\Lambda_0 \gtrsim 10$. The VSI modes start to grow when $\Lambda_0 \la 10$. Fundamental body modes B1 and C1 converge to the hydrodynamic growth rates at $\Lambda_0 \sim 1$. On the other hand, surface modes and high order body modes show slower transitions to their hydrodynamic growth rates, requiring $\Lambda_0 \sim 0.1$.  
The right panel of Figure \ref{fig:p} shows the maximum growth rate as a function of $\beta_{Z0}$ and $\Lambda_0$. Large $\Lambda_0 \sim 10^2$ corresponds to the ideal MHD limit, where maximum growth rate drops to $\sim 5\times10^{-3}$ for $\beta_{Z0}=10^4$. It can be seen that $\Lambda_0 < 1$ is required for the fastest growing modes to recover hydrodynamic results in a wide range of $\beta_{Z0}$ from $10^4$ to $10^{10}$.

The above results are similar to that for toroidal fields, which indicates that the qualitative effect of a magnetic field and resistivity on the VSI does not depend on the background field geometry. 

\subsection{VSI VS MRI}\label{sec:MRI}

\begin{figure*}
\includegraphics[width=0.95\textwidth]{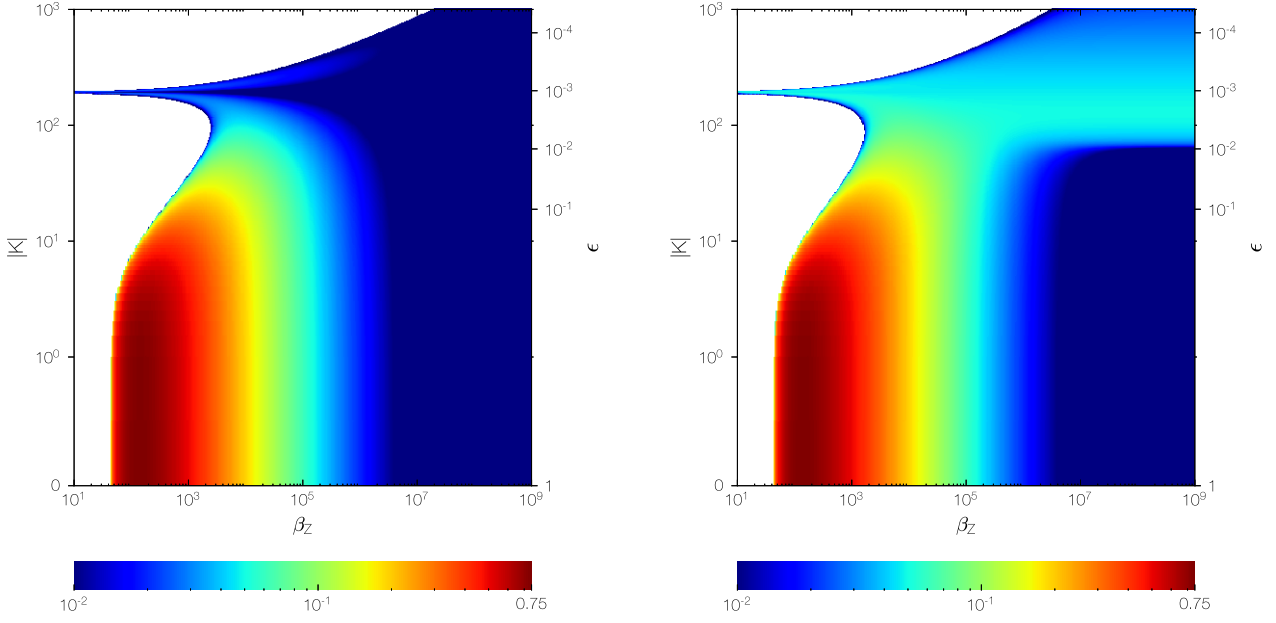}
\caption{Contours of growth rates as functions of $\beta_\rmz$ and radial wavenumber $|K|$ by Equation \eqref{app:drc} in the ideal MHD limit. The vertical wavenumber is set to be $\rmk_\rmz= 2\pi/H$, and $\epsilon=\rmk_\rmz^2/\rmk^2$.  Left:  no vertical shear $R\p \Omega^2/\p Z = 0$. Right: with vertical shear $R\p \Omega^2/\p Z =-0.1$.}
\label{fig:mri}
\end{figure*}

\begin{figure*}
\includegraphics[width=0.95\textwidth]{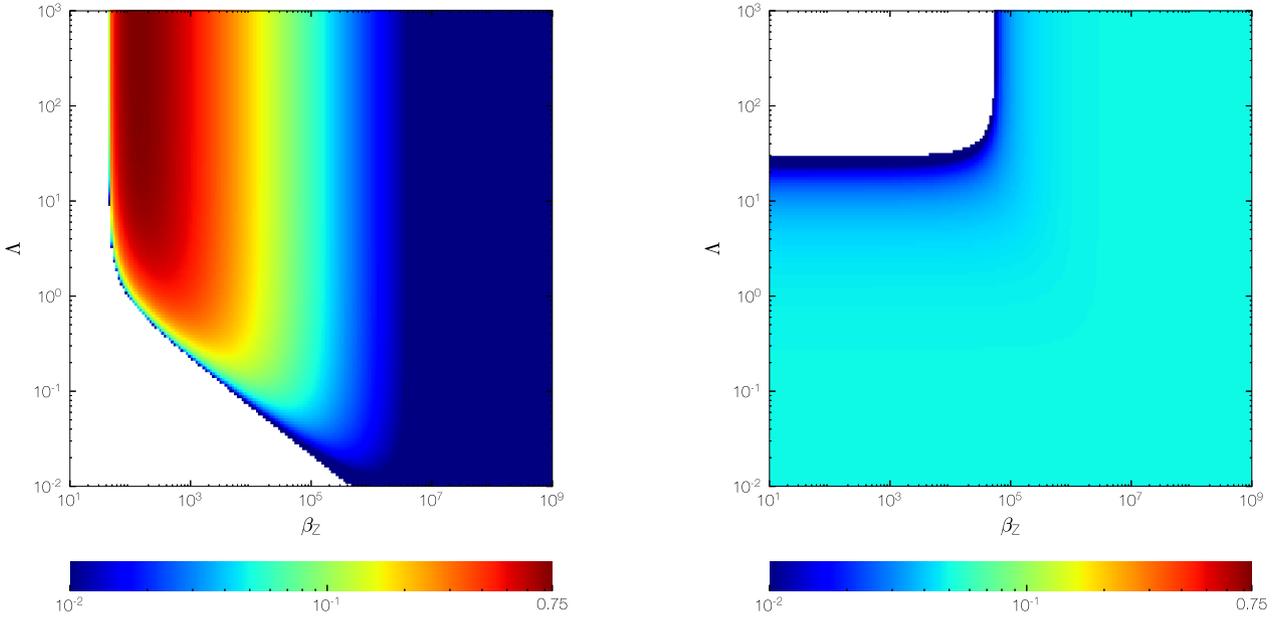}
\caption{Contours of growth rates as functions of $\beta$ and Ohmic Els\"{a}sser number $\Lambda$ by Equation \eqref{app:drc}. The vertical wavenumber and vertical shear are set to be $\rmk_\rmz= 2\pi/H$ and $R\p \Omega^2/\p Z = -0.1$. Left: MRI modes ($|K|=0$). Right: VSI modes ($|K|=150$).}
\label{fig:mrio}
\end{figure*}

\begin{figure}
\includegraphics[width=0.45\textwidth]{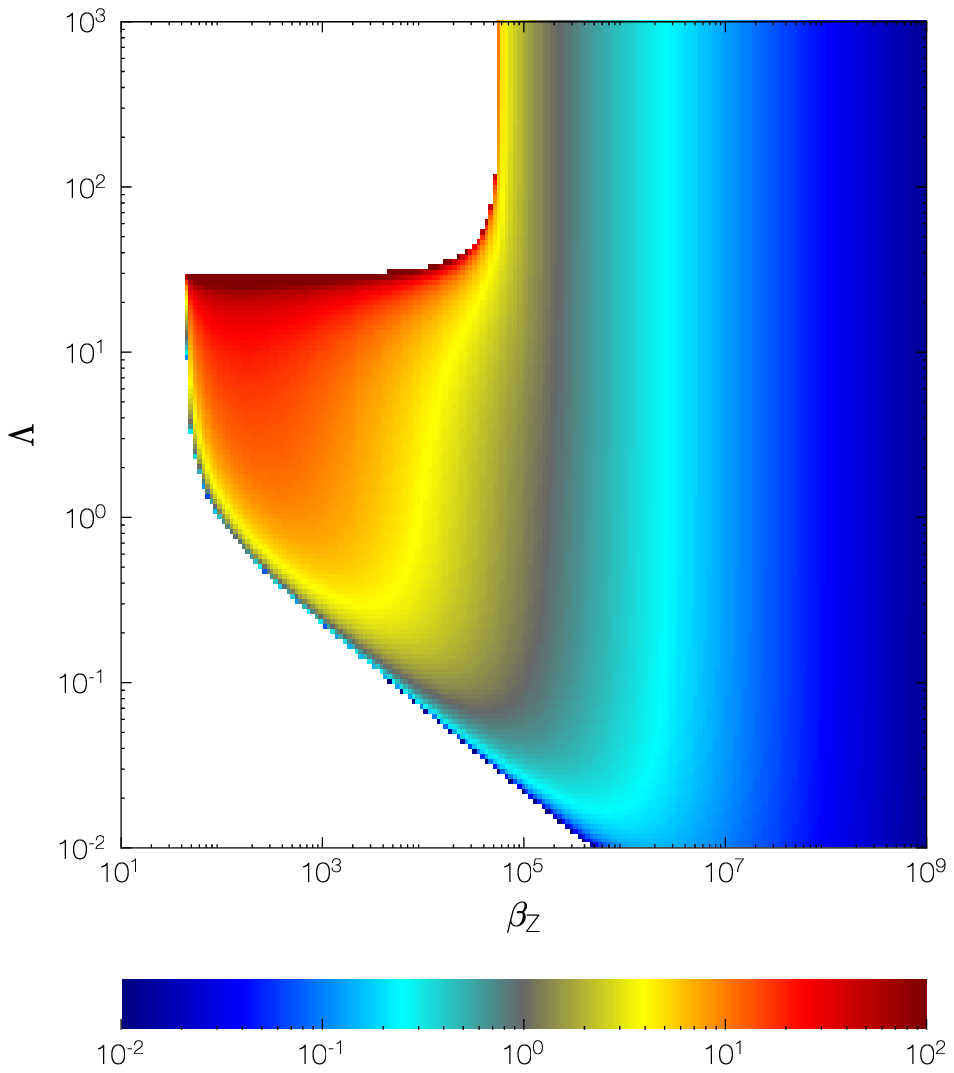}
\caption{The ratio of MRI growth rates to VSI growth rates computed from local dispersion relation Equation \eqref{app:drc}.}
\label{fig:mv}
\end{figure}

To better compare the VSI and the MRI, we perform a vertically local linear analysis in the incompressible limit with Ohmic resistivity. In this local model, background quantities are assumed to be uniform and its values set to that in the vertically global model (\S\ref{sec:gb}) at some fiducial height. We take the vertical shear rate $R\partial\Omega^2/\partial Z$ as an input parameter. We consider axisymmetric perturbations that are proportional to $\exp(\sigma t +\ik_\rmR R+\ik_\rmz Z)$. The wavenumber vector is denoted by $\bk=(k_R,0,k_Z)$.   

In the incompressible limit, the linearized equations derived from Equations \eqref{eq:cont}, \eqref{eq:eom}, and \eqref{eq:idc} become
\begin{align}
\ik_\rmR v_\rmR' + \ik_\rmz v_Z' = 0,
\end{align}
\begin{align}
\sigma \rmv_\rmR' -2\Omega\rmv_\phi' + \ik_\rmR\frac{P'}{\rho} + \frac{\ik_\rmR}{\mu_0\rho} \bB \cdot \bB'  - \frac{\rmi \rmB_\rmR' }{\mu_0\rho} \bk \cdot \bB =0 ,
\end{align} 
\begin{equation}
\sigma \rmv_\phi' + \frac{\kappa^2}{2\Omega}\rmv_\rmR' + \dv{\rmv_\phi}{Z}\rmv_\rmz' - \frac{\rmi  \rmB_\phi' }{\mu_0\rho} \bk \cdot \bB =0    ,
\end{equation} 
\begin{align}
\sigma\rmv_\rmz' + \ik_\rmz\frac{P'}{\rho}  + \frac{\ik_\rmz}{\mu_0\rho} \bB \cdot \bB' - \frac{\rmi \rmB_\rmz' }{\mu_0\rho} \bk \cdot \bB =0   ,
\end{align} 
\begin{equation}
\tilde{\sigma} \rmB_\rmR' - \rmi \bk \cdot \bB \rmv_\rmR'=0   ,
\end{equation} 
\begin{align}
\tilde{\sigma}\rmB_\phi' - \rmi \bk \cdot \bB \rmv_\phi' - \bigg[\frac{\kappa^2}{2\Omega}-2\Omega\bigg] \rmB_\rmR'  -\dv{\rmv_\phi}{Z}\rmB_\rmz' =0   ,
\end{align} 
\begin{equation}
\tilde{\sigma} \rmB_\rmz' - \rmi\bk \cdot \bB \rmv_\rmz'=0   ,
\end{equation} 
where 
\begin{equation}
\rmk^2 = \rmk_\rmR^2+\rmk_\rmz^2, \nn
\end{equation}
 \begin{equation}
 \tilde{\sigma} = \sigma +\eta \rmk^2. \nn
 \end{equation}
The above equations give a dispersion relation 
\begin{align}
[\sigma  \tilde{\sigma } + (\bk \cdot \bva)^2]^2 + \frac{\rmk_\rmz^2}{\rmk^2} ( \kappa^2 &-  A) [\tilde{\sigma}^2 + (\bk \cdot \bva)^2]  \nn  \\
&- 4\Omega^2  (\bk \cdot \bva)^2  \frac{\rmk_\rmz^2}{\rmk^2}  = 0, 
\label{app:drc}
\end{align}
 where
 \begin{equation}
 A = \frac{\rmk_\rmR}{\rmk_\rmz} \frac{R \p\Omega^2}{\p Z}. \nn
 \end{equation}
Equation \eqref{app:drc} generalizes that of LP18 to include Ohmic resistivity. Appendix \ref{app:mri} explores this dispersion relation in more detail in the limit $\left|k_RB_R\right|\ll \left|k_ZB_Z\right|$. Here, we solve Equation \eqref{app:drc} in full to investigate the dominance of MRI and VSI as functions of disk magnetizations $\beta_Z$, radial wavenumbers $K$, and Els\"{a}sser numbers $\Lambda$. A vertical wavenumber $\rmk_Z=2\pi/H$ is fixed in numerical calculations. VSI modes require $k_R$ and $k_Z$ with opposite signs  if $\p\Omega/\p Z < 0$  \citep{nelson_etal13,lp18}, as we will consider below, hence we use $|K|$ to denote the absolute value of $K$. Other background quantities are evaluated at $Z=2H$.

The left panel of Figure \ref{fig:mri} shows local growth rates of MRI modes in the ideal MHD limit ($\eta=0$) without vertical shear ($R\p \Omega^2/\p Z = 0$). In the white regions, the MRI is quenched by strong magnetizations. It can be seen that the MRI modes prefer small $|K|$, with a maximum growth rate of $s=0.75$ at $\beta_\rmz =137$.
In the right panel of Figure \ref{fig:mri}, we show growth rates of VSI modes by setting $R\p \Omega^2/\p Z = -0.1$. The fastest growing VSI modes prefer weak magnetization and large $|K|$, in contrast to fast growing MRI modes. Local VSI modes with $K=150$ and $\beta_\rmz=10^9$ have a growth rate $s = 0.05$, which is less than that of the fastest growing surface mode $s = 0.094$ obtained from vertically global analysis shown in the right panel of Figure \ref{fig:ideal_mode}. 

We next consider resistive disks. In the left panel of Figure \ref{fig:mrio}, we show growth rates as functions of $\beta_\rmz$ and $\Lambda$ for MRI modes by setting $K=0$. In contrast to the VSI, MRI growth rates decline towards small $\Lambda$ as the MRI is dampened. In the right panel of Figure \ref{fig:mrio}, we show growth rates of mostly VSI modes by setting $K=150$. For $\beta_Z \gtrsim 10^5$, we obtain VSI growth rates in the hydrodynamic limit. For $\beta_Z \lesssim 10^5$, the VSI is dampened for $\Lambda \gtrsim 10$, but revived with small $\Lambda$ or strong resistivity. 

Finally, in Figure \ref{fig:mv}, we show the ratio of MRI growth rates obtained from the left panel of Figure \ref{fig:mrio}, to VSI growth rates obtained from the right panel of Figure \ref{fig:mrio}. For $\beta_Z\gtrsim 10^5$ the VSI dominates over the MRI. For $\beta_Z\lesssim 10^5$, which includes PPDs with typical $\beta_Z\sim 10^4$, the VSI dominates if $\Lambda \lesssim 0.09$. 

\section{Discussion}\label{sec:di}
\subsection{Comparison with previous works}

LP18 carried out local linear analyses of the VSI in the ideal MHD limit with an exact background equilibrium solution for a purely poloidal field. In this work, we set up global equilibria for a purely azimuthal field. Comparing our numerical solutions to local analytical results, we find that the overall mode behaviour that magnetization can stabilize the VSI is in agreement with each other. Global numerical simulations of magnetized PPDs carried out by \cite{cb20} indeed show that magnetism tends to suppress the VSI growth, whereas ambipolar diffusion acts to revive VSI modes. Their simulations also show the absence of surface modes, which is consistent with our findings that surface modes are the first to be dampened with increasing field strengths. 

Global simulations of PPDs initially threaded by a large-scale poloidal magnetic field suggest a field configuration dominated by the toroidal component once the disk reaches a quasi-steady state (e.g. \citealp{bai17,bethune_etal17,cb20}). This suggests that, as far as the VSI is concerned, it is the disk model with an azimuthal field that is more relevant, as employed in most of this paper. Furthermore, the magnetization in our toroidal field model is parametrized by $\beta$, which does not depend on the orientation of the magnetic field. Our results are thus applicable to the aforementioned simulations wherein the toroidal field reverses polarity across the disk midplane because of the Keplerian shear. 

\subsection{Application to PPDs}\label{sec:app}

In the outer part of the PPDs ($\gtrsim 30$ AU), ambipolar diffusion is the dominant non-ideal MHD effect, with Els\"{a}sser numbers approximately unity. For a purely azimuthal field, ambipolar diffusion acts as an effective resistivity with field dependency (see \S\ref{sec:toraxis}). Hence, our results for Ohmic resistivity are also applicable to ambipolar diffusion. A value of $\Lambda_0=1$ and $\beta_{\phi0}=10^2$ gives a maximum growth rate of $s = 0.087$ (Figure \ref{fig:ohmic}), which is close to the hydrodynamic result, $s = 0.094\OmK$. In the inner part of the PPDs, Ohmic resistivity becomes the dominant non-ideal MHD effect, though the Hall effect also contribute. At 2 AU, the midplane $\Lambda_0=5\times10^{-4}$ \citep{bai17} gives a maximum growth rate of $s = 0.094$ for a wide range from $\beta_{\phi0}=10$ to $\beta_{\phi0}=10^9$ (Figure \ref{fig:ohmic}), as a small $\Lambda_0$ enables the recovery of hydrodynamic results. 

Our locally isothermal disk models, which correspond to instantaneous cooling, favor the VSI because there is no stabilizing effect from vertical gas buoyancy \citep[N13,][]{ly15}. However,
in a realistic PPD cooling timescales are finite and is sensitive to stellar irradiation and dust properties  \citep{malygin_etal17,pk20,flock_etal20}. 
In magnetized disks, the magnetic field will provide extra stabilization via magnetic buoyancy in addition to gas buoyancy. Thus, we expect that in PPDs the required cooling time may be shorter than that estimated based on purely hydrodynamic models \citep{ly15}, unless $\Lambda_0$ is small enough to diminish the stabilizing effect from magnetic fields. Detailed analyses should be conducted to give new critical cooling timescales in magnetized disks.

The analysis we present with Ohmic resistivity points to future directions in including ambipolar diffusion and the Hall effect, the latter of which requires a poloidal field. A few obstacles and complications needs to be resolved when incorporating these two non-ideal MHD effects. Firstly, it is difficult to find appropriate background equilibria in a global model because of the vertical shear, especially for in presence of poloidal magnetic fields \citep{ogilvie97}.
Furthermore, ambipolar diffusion gives rise to anisotropic damping and introduces the ambipolar shear instability \citep{bb94,kb04,kunz08}. The Hall effect will further introduce the Hall shear instability, and its effect is polarity dependent \citep{bt01,kunz08}. All of these effects will complicate the problem and deserve step-by-step analyses in local and global disk models in the future. 

\subsection{Implications to dust dynamics}

Small dust grains tend to settle towards the disk midplane \citep{dubrulle_etal95}. However, the VSI drives turbulence that can vertically mix up dust particles \citep{sk16,flock_etal17,flock_etal20}. On the other hand, our results show that strong magnetization can stabilize the VSI away from the disk midplane, implying a limit on the vertical extent of the ensuing VSI turbulence and therefore a maximum dust layer thickness, $H_\mathrm{d,max}$. For definiteness, consider a purely toroidal field with constant $\beta$ and neglect non-ideal MHD effects. The destabilizing vertical shear, $R\p\Omega^2/\p Z$, competes against the stabilizing magnetic buoyancy, $N_Z^2$. We therefore expect the VSI to be suppressed where $\left|R\p\Omega^2/\p Z\right|/N_Z^2 < \zeta$, where $\zeta$ is some critical ratio\footnote{For the hydrodynamic VSI, $\zeta \sim \Omega t_\mathrm{cool}$, where $t_\mathrm{cool}$ is the cooling timescale \citep{ly15}.}. Using Equations \eqref{eq:shear} and \eqref{eq:Nzbeta}, we estimate 
\begin{align}
\frac{H_\mathrm{d,max}}{H} = \frac{\beta h \left|q_T\right|}{\zeta},
\label{Hdmax_beta}
\end{align}
assuming $\beta\gg 1$ and a thin disk. The example in Figure \ref{fig:igenf} with $\beta=10^2$, $h=0.05$, and $q_T=1$ show that gas motions are negligible for $|Z|\gtrsim 2H$, which suggest $\zeta\simeq 2.5$.

One may ask if $H_\mathrm{d,max}$ can be made sufficiently small to constrain particles to a dense midplane layer that can undergo, for example, the streaming instability and hence facilitate planetesimal formation \citep{youdin05,johansen09}. For dynamical growth, the streaming instability requires a local dust-to-gas mass ratio $\nu \gtrsim 1$. The metallicity is $\Sigma_\mathrm{d}/\Sigma_\mathrm{g} = \nu H_\mathrm{d}/H$, where $\Sigma_\mathrm{d}$ and $\Sigma_\mathrm{g}$ are the dust and gas surface densities, respectively, and $H_\mathrm{d}$ is the characteristic dust scale height. Thus, if we take $H_\mathrm{d} = H_\mathrm{d,\max}/2$, then 
\begin{align}
\beta = \frac{2\zeta}{\nu  \left|q_T\right|h}\frac{\Sigma_\mathrm{d}}{\Sigma_\mathrm{g}}.  
\end{align}
Inserting typical PPD values  $\left(q_T,\,h,\,\Sigma_\mathrm{d}/\Sigma_\mathrm{g}\right)\simeq\left(1, \, 0.05,\,0.01\right)$, we find an equipartition field strength, $\beta=1$, would be required to confine dust into a thin layer such that $\nu \sim 1$ by quenching the VSI elsewhere.
Such a strong field is unrealistic for PPDs (e.g. \citealp{simon_etal13a,bai15}). This suggests that magnetic fields do not affect the vertical dust structure in PPDs through its geometric effect on the VSI. 

Instead, magnetic effects likely manifest through weakening the VSI and hence the ensuing turbulence, as found in this work and \cite{cb20}. This determines $H_\mathrm{d} \simeq \sqrt{\delta_Z/\mathrm{St}}H_\mathrm{g}$ \citep{dubrulle_etal95}, where $\mathrm{St}$ is the particle Stokes number and $\delta_Z$ is the dimensionless vertical diffusion coefficient associated with VSI turbulence. We expect $\delta_Z$ to drop with larger $\beta$ and $\delta_Z$ to increase with non-ideal MHD effects. This relation should be calibrated with future simulations, which can then be used to estimate magnetic field strengths from the vertical distribution of dust in PPDs. 

\section{Conclusions}\label{sec:sum}

In this work, we perform linear analyses of the VSI under the ideal MHD limit and with Ohmic resistivity. A vertically global and radially local disk model is employed to properly accommodate the characteristic VSI modes of elongated vertical wavelengths. A locally isothermal equation of state is assumed to better focus on the effect of magnetism. Our main findings are summarized as follows. 

\begin{itemize}
\item In the ideal MHD limit, magnetic fields operate as a stabilizing effect to suppress the growth of VSI modes.
Surface modes are the first to vanish rather than body modes with increasing magnetic field strengths.
Ohmic resistivity acts as destabilizing effect to assist the VSI growth. 
\item In weakly magnetized disks, surface modes show maximum growth rates at large radial wavenumbers, while in strongly magnetized disks, surface modes are dampened, while body modes are dominant and prefer intermediate radial wavenumbers. Large disk aspect ratios or vertical shear rates leads to fast VSI growth. 
\item The MRI modes appear when a poloidal magnetic field is present. In the local analysis, we find that MRI and VSI modes dominate at different $\beta_\rmz$ and $\Lambda$ in the ideal MHD limit. MRI prefers relatively strong disk magnetizations and small radial wavenumbers. The VSI modes are most effective at weak magnetizations and large radial wavenumbers. With Ohmic resistivity, a typical value of $\beta_\rmz=10^4$ in PPDs results in a critical $\Lambda \lesssim 0.09$ for the dominance of the VSI. 
\end{itemize}

\section*{Acknowledgements}

We are pleased to thank Xue-Ning Bai, Henrik Latter, and Gordon Ogilvie for fruitful discussions. CC acknowledges the support from Department of Applied Mathematics and Theoretical Physics at University of Cambridge. 
MKL is supported by the Ministry of Science and Technology of Taiwan under grant 107-2112-M-001-043-MY3 and an Academia Sinica Career Development Award (AS-CDA-110-M06). 

\addcontentsline{toc}{section}{Acknowledgements}

Software: \href{https://dedalus-project.org/}{DEDALUS} \citep{burns_etal20} and \href{https://github.com/DedalusProject/eigentools}{EIGENTOOLS} package.

\section*{Data Availability}

The data underlying this article will be shared on reasonable request to the corresponding author.

\appendix
\section{Properties of the Local Dispersion Relation}\label{app:mri}

We explore the properties of the local  dispersion relation Equation \eqref{app:drc} in the limit of $\left|k_RB_R\right|\ll \left|k_ZB_Z\right|$. When there is no vertical shear, $R \p \Omega^2 /\p Z =0$, Equation \eqref{app:drc} is identical to the dispersion relation for the MRI in Equation (22) of \citet{sano99}. When there is no magnetic field, $\rmv_{\az}=0$ and $\eta=0$, our dispersion relation recovers Equation (34) of \citet{gs67} in the case when Brunt-V\"ais\"al\"a frequency vanishes. When there is a magnetic field but $\eta=0$, the dispersion relation recovers Equation (58) of LP18.

In the hydrodynamic limit, $\rmv_{\az}=0$ and $\eta=0$, the dispersion relation resembles that of pure VSI,
\begin{equation}
\sigma^2 +  \frac{\rmk_\rmz^2}{\rmk^2} (\Omega^2 - A) =0.
\label{local_pure_vsi}
\end{equation}
Defining $\epsilon \equiv k_Z^2/k^2$ and taking $\p\Omega/\p Z < 0$ without loss of generality, we can write
$\rmk_\rmR/\rmk_\rmz = - (1/\epsilon-1)^{1/2}$. The most unstable wavenumbers satisfy
\begin{equation}
\Omega = -\frac{2\epsilon(1/\epsilon -1) -1}{\epsilon(1/\epsilon-1)^{1/2}}\frac{R\p \Omega}{\p Z},
\end{equation}
The vertical shear rate is on the order of $-R \p \Omega /\p Z \sim O(\Omega h)$ via Equation \eqref{eq:dO2dZ2}, hence $-\Omega/(R \p \Omega /\p Z) \sim O(h^{-1})$, which requires $\epsilon$ to be small. Therefore, the above equation can be written as 
\begin{equation}
h^{-1} \approx \lim_{\epsilon \to 0} \frac{2\epsilon(1/\epsilon -1) -1}{\epsilon(1/\epsilon-1)^{1/2}}  \approx \epsilon^{-1/2},
\end{equation}
so that $\rmk_\rmz^2/\rmk^2 \sim h^2$ and the maximum growth rate is $\sigma_{\mathrm{max}} \sim O(\Omega h)$, which recovers the results in N13.

\begin{figure*}
\includegraphics[width=0.9\textwidth]{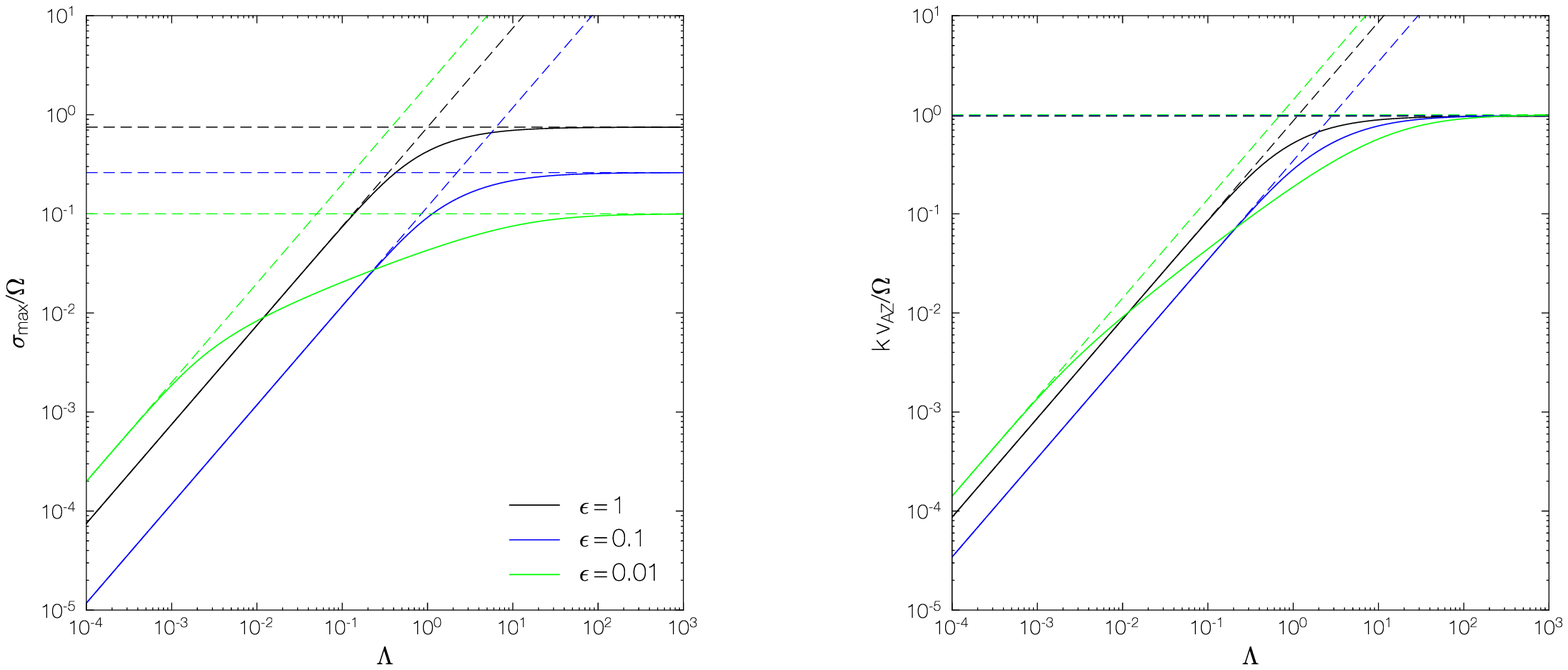}
\caption{Maximum growth rates and most unstable wavenumber are shown as functions of Ohmic Els\"{a}sser number $\Lambda$ by \ref{app:maxk} at fixed Alfv\'en velocity and $R \p \Omega^2/\p Z = -0.1$.  Dashed lines are asymptotic solutions in the limit of $\eta \to 0$ (\ref{app:w1} and \ref{app:k1}) and $\eta \to \infty$ (\ref{app:w2} and \ref{app:k2}). }   
\label{fig:app}
\end{figure*}

The full dispersion relation \eqref{app:drc} is implicit in $\sigma$ and $\rmk$. To obtain the maximum growth rate, we take the derivative $\p/\p \rmk$ for each term on the left-hand side and assume $\epsilon$ to be a constant. This yields a relation between the maximum growth rate $\sigma_{\mathrm{max}} $ and the most unstable wavenumber $\rmk$,
\begin{equation}
2\rmk^2 = \frac{ \rmv_\az^2(-2\sigma^2\epsilon+3\Omega^2\epsilon^2+A\epsilon^2) - 2\sigma\eta[\sigma^2+(\Omega^2-A)\epsilon]}{ \rmv_\az^4\epsilon^2+2\eta\sigma\rmv_\az^2\epsilon + \eta^2[\sigma^2 + (\Omega^2-A)\epsilon]}.
\label{app:maxk}
\end{equation}
The maximum growth rate is computed numerically by substituting $\rmk$ in Equation \eqref{app:drc}, and the results are shown in Figure \ref{fig:app}. In the limit of weak resistivity $\eta \to 0$, the maximum growth rate and the most unstable wavenumber are
\begin{equation}
\sigma^2_{\mathrm{max}} = \frac{9\Omega^2 + A[(6\Omega^2+A)/\Omega^2]}{16} \epsilon, 
\label{app:w1}
\end{equation}
\begin{equation}
\rmk^2 = \frac{15\Omega^2 + A(2\Omega^2-A)/\Omega^2}{16\rmv_{\az}^2}.
\label{app:k1}
\end{equation}
When there is no vertical shear $A=0$, these expressions recover the MRI channel modes.
When $\eta \to \infty$,
\begin{equation}
\sigma^2_{\mathrm{max}} = \frac{(3\Omega^2+A)^2}{16(\Omega^2-A)^2} \frac{\rmv_{\az}^4}{\eta^2} \epsilon^2 , 
\label{app:w2}
\end{equation}
\begin{equation}
\rmk^2 = \frac{(3\Omega^2+A)}{4(\Omega^2-A)}\frac{\rmv_{\az}^2}{\eta^2}  \epsilon,
\label{app:k2}
\end{equation}
and we recover \citet{sano99} when $A=0$. 

\bsp
\label{lastpage}

\bibliographystyle{mnras}
\bibliography{vsi}

\end{document}